\documentclass[11pt,a4paper,preprintnumbers,nofootinbib]{revtex4}
\pdfoutput=1

\usepackage[T1]{fontenc}
\usepackage[utf8]{inputenc}

\usepackage[english]{babel}

\usepackage{graphicx}

\usepackage{amsmath}
\usepackage{amsfonts}
\usepackage{amssymb}
\usepackage{amstext}
\usepackage{isomath}

\usepackage[strict, separate-uncertainty, sticky-per, exponent-product=\cdot]{siunitx}

\usepackage{microtype}

\usepackage{multirow}

\usepackage{hyperref}
\usepackage{placeins}
\usepackage{verbatim}
\usepackage{xcolor}

\begin{document}

\title{Leptoquark Flavor Patterns \& $B$ Decay Anomalies}

\author{Gudrun Hiller}
\email{ghiller@physik.uni-dortmund.de}
\author{Dennis Loose}
\email{dennis.loose@udo.edu}
\author{Kay Sch\"onwald}
\email{kay.schoenwald@udo.edu}
\affiliation{Fakult\"at Physik, TU Dortmund, Otto-Hahn-Str.4, D-44221 Dortmund, Germany}
\preprint{DO-TH 16/17}

\begin{abstract}
Flavor symmetries that  explain masses and mixings of the standard model fermions dictate  flavor patterns for the couplings of scalar and vector leptoquarks  to the
standard model fermions. 
A generic feature is that couplings to $SU(2)$-doublet leptons are  suppressed at least by one spurion of the discrete non-abelian symmetry breaking, responsible for neutrino mixing, while couplings to charged lepton singlets can be order one.
We obtain  testable patterns  including those that predominantly couple to 
a single lepton flavor, or two, or  in a skewed way. They induce lepton non-universality, which we contrast
to current anomalies in $B$-decays.
We find maximal effects in $R_{D}$ and $R_{D^*}$ at the level of  $\sim$10  percent  and few percent, respectively, while
leptoquark effects in $R_{K^{(*)}}$ can reach  order  few$\times 10$ percent.
Predictions  for  charm and kaon decays and $\mu-e$ conversion are worked out.

\end{abstract}

\maketitle

\section{Introduction}

Generational structure  and its manifestation through  fermion masses and mixings is one of the key ingredients  of the Standard Model (SM), however,
the  origin of flavor remains a puzzle. While quark flavor solely 
resides in the Yukawa couplings to the Higgs, flavor breaking in the lepton sector is even less clear-cut  as neutrino masses could stem from a different mechanism.

Flavor symmetries  explain the observed masses and mixings of the SM fermions, however,  viable solutions are not unambiguous.
 Physics beyond the SM provides opportunities for new insights,
as new couplings allow for different combinations of flavor symmetry breaking. A well-known example is the minimal supersymmetric SM. Here
masses and mixings of scalar quarks and leptons are present which allow to probe also non-chiral combinations  of matter bilinears \cite{Nir:1993mx}.

Here we consider  scalar and vector leptoquark extensions of the SM \cite{Buchmuller:1986zs,Davidson:1993qk}.
Representations for $SU(3)_C \times SU(2)_L \times U(1)_Y$ and  renormalizable couplings to SM fermions are given in Appendix \ref{app:L}.  We assume that proton decay is safe, note, however, to ensure this
 requires further model-building for some of the leptoquarks \cite{Dorsner:2012nq}.
There are in total twelve
different types of  flavor-matrices  that appear in leptoquark models with couplings to
$SU(2)_L$ doublet quarks $Q$ and leptons $L$, and $SU(2)_L$ singlet quarks $U,D$ and charged leptons $E$, schematically,
\begin{align} \label{eq:YAB}
Y_{AB}, Y_{\bar A B} \, , \quad A=Q,U,D \,, ~~ B=L,E \, ,
\end{align}
where rows and columns correspond to quarks and leptons, respectively. To simplify the discussion, in the following these couplings are denoted Yukawa matrices for both scalar and vector leptoquarks.

Table \ref{tab:models} shows for which leptoquark scenario  which type of yukawa is present. Also indicated is  by a checkmark which yukawa contributes at tree level to (semi-)leptonic  flavor changing neutral current 
(FCNC) transitions in the down and up-quark sector.
Contributions to charged currents, {\it e.g.}, $b \to c \ell \nu$  in chirality-preserving  four-fermion interactions, $\bar Q \gamma_\mu (\sigma^a) Q \bar L \gamma^\mu (\sigma^a)L$
\cite{Bhattacharya:2014wla} arise  from  $Y_{QL}$, $Y_{\bar QL}$ only. $\sigma^a$ denote the Pauli-matrices.
$S_{1,2}$ and $V_{1,2}$ induce charged currents through a combination of their two couplings present, resulting in chirality-flipping operators.

\begin{table}
    \centering
    \begin{tabular}{c||c|c|c|c|c|c||c|c|c|c|c|c}
       $AB$ & $QL$ & $\bar QL$ &  $UL$ &  $\bar UL$ &  $DL$ & $\bar DL$ &  $QE$ & $\bar QE$ & $UE$ &  $\bar UE$ &  $DE$ &  $\bar DE$ \\ \hline 
        model & $S_{1,3}$ &  $V_{1,3}$ &   $ \tilde V_2$ & $S_2$ & $V_2$ &  $ \tilde S_2$ & $V_2$ &  $S_2$ & $S_1 $ & $\tilde V_1$ & $\tilde S_1$ & $V_1$ \\
        down-type FCNC & \checkmark  & \checkmark  & --  & --  & \checkmark  & \checkmark & \checkmark  & \checkmark  &-- & -- & \checkmark  & \checkmark \\
       up-type  FCNC & \checkmark & \checkmark & \checkmark & \checkmark & --  & --& \checkmark & \checkmark & \checkmark & \checkmark & -- & -- \\
    \end{tabular}
    \caption{ Leptoquark couplings $Y_{AB}$ and $Y_{\bar AB}$ as they appear  in various leptoquark models as well as in tree level down-type  quark FCNCs and up-type quark FCNCs.
    Models $S_{1,2}$ and $V_{1,2}$ have two yukawas each.}    
    \label{tab:models}   
\end{table}

Patterns based on  a $U(1)_\text{FN}$-Froggatt--Nielsen (FN) symmetry  \cite{Froggatt:1978nt}, combined with a non-abelian discrete symmetry, $A_4$, have been worked out previously for leptoquarks coupling to  lepton doublets \cite{Varzielas:2015iva,deBoer:2015boa}. Here we provide further details and
 flavor patterns involving singlet leptons. 
The $U(1)_\text{FN}$ explains hierarchies in the quark sector and for the charged lepton masses, while non-abelian discrete subgroups of $SU(3)$ can accommodate 
neutrino mixing \cite{Barger:2012pxa}. We stress that leptoquark extensions of the SM are special as they can access both quark and lepton flavor.

Leptoquark models have  been considered recently  in the context of lepton non-universality (LNU) and lepton flavor violating (LFV) observables in  semileptonic $b \to s$ and
$b \to c$  decays.
While there is vast literature on leptoquark models addressing LNU in $B \to D^{(*)} \ell \nu$ decays, the leptonic flavor structure  in these studies is either of an assumed, simple type such as third generation only,
of within minimal flavor violation and variants thereof, or parametrical amended by experimental constraints \cite{Sakaki:2013bfa,Dorsner:2013tla,Alonso:2015sja,Calibbi:2015kma,Barbieri:2015yvd,Freytsis:2015qca,Bauer:2015knc,Fajfer:2015ycq,Becirevic:2016yqi,Sahoo:2016pet}, and references therein.
Our aim is here to close this gap and work out leptoquark effects based on flavor symmetries.
For previous leptoquark studies regarding LNU in $B \to K \ell \ell$ decays, see, for instance,
\cite{Hiller:2014yaa,Bauer:2015knc,Sahoo:2015wya,Varzielas:2015iva,Alonso:2015sja,Becirevic:2015asa,Gripaios:2014tna,Fajfer:2015ycq,Freytsis:2015qca,Pas:2015hca,Becirevic:2016yqi}.

The plan of the paper is as follows:
In Sections \ref{sec:patterns} and \ref{sec:A4}  we summarize the main model-building tools for obtaining flavor patterns for Yukawa matrices, and present patterns in
Section \ref{sec:A4}.
In Section \ref{sec:phen} we work out  phenomenological implications for  $B \to D^{(*)}  \ell \nu$, $B \to K^{(*)} \ell \ell$, rare charm and $K$ decays and $\mu-e$ conversion.
We conclude in Section \ref{sec:con}. Auxiliary formulae and tables are given in Appendices \ref{app:L}-\ref{sec:cu}.

\section{Flavor Hierarchies \label{sec:patterns}}

We discuss the hierarchical  structure  of the leptoquark couplings due to an $U(1)_\text{FN}$ which explains flavor in the quark sector and the masses of the charged leptons. To obtain mixing in the lepton sector we additionally  invoke a discrete non-abelian symmetry, $A_4$ \cite{Altarelli:2010gt}, discussed in Section \ref{sec:A4}.

 FN-charges $q$ for quarks and leptons are obtained in \cite{Chankowski:2005qp}; for instance, a realistic  set is given by
 \begin{align} \label{eq:charges}
 q(\bar Q)&=q(U)=(4,2,0)  \, , \quad q(D)=(3,2,2) \,   , \quad q(E)=(4,2,0) \, ,  \quad q(L) =0  \, .
 \end{align}
 By choosing this as our  benchmark, we assumed that the Higgs vacuum expectation values (VEVs) responsible for up- and down-quark masses are similar in size to cover a SM-like situation in which
 they are identical. In supersymmetric variants or multi-Higgs models with larger values of $\tan \beta$, the Higgs VEV ratio, 
 smaller values of  $q(D)$ are possible.
Corresponding effects in the lepton sector can be taken into account by adjusting the VEV of the non-abelian flavor symmetry breaking responsible for charged lepton masses.
 
The parametric suppression of the leptoquark yukawas  in terms of powers of $\lambda \sim 0.2$ is then determined as
 \begin{align} \label{eq:yab}
 \left ( Y_{AB} \right)_{ij}\sim \lambda^{| q(A_i) +q(B_j)|} \, , \quad  \left ( Y_{\bar AB} \right)_{ij}\sim \lambda^{|-q(A_i) +q(B_j)|}  \, ,
 \end{align}
 where we assumed that  the leptoquarks are uncharged under the FN-symmetry. Allowing for a finite charge would rescale the overall
 size of the yukawas.
 
 {}From Eq.~(\ref{eq:yab}) 
it is apparent that, unlike in the SM,  charges from the quark sector can interfere with the ones from the lepton sector.
In particular,  with assignments Eq.~(\ref{eq:charges}) 
 cancellations can arise for  $QE, \bar D E$  and $\bar U E$ corresponding to the vector leptoquark scenarios
 $V_2, V_1$ and $\tilde V_1$, respectively. This causes the hierarchy expected from the quark masses to be inverted  in these scenarios.
 For  $\bar QE, D E$  and $U E$ corresponding to the scalar  leptoquark scenarios
 $S_2, \tilde S_1$ and $S_1$, respectively,  the hierarchies will be stronger than from the quarks alone.
 If instead the singlet-lepton charges would be chosen with opposite sign to the quark ones, the effects would swap, that is, an inversion of hierarchies would occur for scalar and an
 increase of hierarchies for vector leptoquarks. Lepton-doublet scenarios would also be  affected in models with $q(L) \neq 0$.
 For $q(L) = 0$,  $Y_{AL} =Y_{\bar A L}$.
 
The breaking of the $U(1)_\text{FN}$ can lead to BSM scalars (flavons) in reach of present or future colliders, with corresponding phenomenology driven by the FN-charges, {\it e.g.,}
\cite{Tsumura:2009yf,Bauer:2016rxs}. 
 Such analysis is interesting, however, beyond the scope of our paper, which focusses on leptoquark-induced BSM effects.

\section{Flavor selection with discrete symmetries \label{sec:A4}}

We employ  the discrete symmetry $A_4\times Z_3$ to model the lepton mixing based on a modification \cite{Varzielas:2012ai} of the original model \cite{Altarelli:2010gt}, which introduces an additional field to account for a non-vanishing value of $\theta_{13}$.
Table \ref{tab:A4Z3_leptons} summarizes the charge assignments of the leptons and the flavon fields, adopted from \cite{Varzielas:2015iva}.
The FN-spurion is uncharged under $A_4\times Z_3$.
\begin{table}[h]
 	\begin{tabular}{c||c|ccc|cc|cc}
		& $L$ & $e_\text R$ & $\mu_\text R$ & $\tau_\text R$ & $\phi_\ell$ & $\phi_\nu$ & $\xi$ & $\xi^\prime$ \\
		\hline
		$A_4$ & 3 & 1 & $1^{\prime}$ & $1^{\prime\prime}$ & 3 & 3 & 1 & $1^\prime$ \\
		$Z_3$ & 1 & 1 & 1 & 1 & 0 & 2 & 2 & 2 \\
	\end{tabular}
    \caption{Non-trivial $A_4\times Z_3$ charge assignments. For leptoquarks, see text. }
    \label{tab:A4Z3_leptons}
\end{table}
The VEVs of the flavons are given as $\langle \phi_\ell \rangle/\Lambda = c_\ell (1,0,0)$, $\langle \phi_\nu \rangle /\Lambda = c_\nu (1,1,1)$ and $\langle \xi^{(\prime)} \rangle/\Lambda = \kappa^{(\prime)}$, where $\Lambda$ denotes a new physics scale related to $A_4$-breaking.
The values of the VEVs are, in general, model-dependent. Typically, $c_{\ell , \nu}, \kappa^{(\prime)}\lesssim\lambda^{\rm few}$ to explain charged lepton and neutrino parameters.
Here and in the following we use the term ``VEV'' for $c_{\ell , \nu}, \kappa^{(\prime)}$ as well. While the latter are dimensionless numbers it should be clear that they do 
not correspond to
renormalizable couplings of the full Lagrangian.

For completeness, we briefly  summarize the multiplication rules for $A_4$. For further 
 information on the basis and group theory of $A_4$ see \cite{Varzielas:2010mp}.
The group has three singlet representations $1$, $1'$, and $1''$ which form a $Z_3$ subgroup with the usual multiplication rules.
Additionally,  $A_4$ has a triplet representation.
Denoting two triplets as $A=(a_1,a_2,a_3)$ and $B=(b_1,b_2,b_3)$ the product reads
\begin{align}   
    \left( A B \right)_{1\phantom{''}} &= a_1 b_1 + a_2 b_3  + a_3 b_2 \sim 1 ,\\
    \left( A B \right)_{1'\phantom{'}} &= a_1 b_2 + a_2 b_1  + a_3 b_3 \sim 1' ,\\
    \left( A B \right)_{1''} &= a_1 b_3 + a_2 b_2  + a_3 b_1 \sim 1'' 
\end{align}
and 
\begin{align}
    \left( AB \right)_s &= \frac{1}{3} \begin{pmatrix} 2a_1b_1 - a_2b_3 - a_3b_2 \\ 2a_3b_3-a_1b_2-a_2b_1 \\ 2a_2b_2-a_3b_1-a_1b_3 \end{pmatrix}, & \left( AB \right)_a &= \frac{1}{2} \begin{pmatrix} a_2b_3-a_3b_2 \\ a_1b_2-a_2b_1 \\ a_3b_1-a_1b_3 \end{pmatrix} ,
\end{align}
with a symmetric ($s$) and an antisymmetric ($a$) triplet. 
    
Firstly, all quarks are considered  $A_4$ singlets (Section \ref{sec:trivial}). In  Section \ref{sec:nontrivial} we discuss patterns for 
individual quark generations in non-trivial singlet-representations of $A_4$.

\subsection{Quarks trivial under $A_4$ \label{sec:trivial}}

\subsubsection{Flavor patterns}

If all quarks have a trivial $A_4$-charge and identical $Z_3$-charges, one needs to distinguish only  between the couplings to right-handed leptons  $Y_{AE} (Y_{\bar AE})$ and  to left-handed leptons $Y_{AL} (Y_{\bar AL})$.
The column structure of the patterns is governed by the $A_4$ and $Z_3$-charges of the leptoquark.
The $A_4$-charge determines to which lepton generation(s) the leptoquark can couple, while the $Z_3$-charge selects the flavon field that mediates the coupling.
We denote the leptoquarks $S_i^\dagger, V_i^\dagger$ generically by $\Delta$ and the charge assignments by $[\Delta]_{A_4}$ etc.

For the left-handed couplings the crucial flavons are the $A_4$-triplets $\phi_\ell$ and $\phi_\nu$, which produce patterns that either isolate a single lepton generation or couple equally to all generations \cite{Varzielas:2015iva}.

For the right-handed leptons, 
terms of the form $A\Delta E$ are $Z_3$-invariant without any additional flavon insertion  for $[\Delta]_{Z_3}=2$. 
(Here and in the following, as in Eq.~(\ref{eq:YAB}), $A$  ($B$) generically denotes quark (lepton) fields.)
In this case one  isolates a single lepton generation, depending on the leptoquark's $A_4$-representation.
Additionally, and in contrast to the respective pattern for the left-handed coupling, the isolated column is suppressed by powers of $\lambda$ due to the FN-charges of the right-handed leptons.
For $[\Delta]_{Z_3}=0$ one additional flavon, $\xi$ or $\xi^\prime$, is needed.
Since the latter  have identical  $Z_3$-charge  but transform under different singlet representations of $A_4$, two lepton generations are isolated.

To  summarize these findings we  introduce the following lepton flavor isolation textures
\begin{align}
    k_e = \begin{pmatrix} * & 0 & 0 \\ * & 0 & 0 \\ * & 0 & 0 \end{pmatrix},\quad
    k_\mu = \begin{pmatrix} 0 & *& 0 \\ 0 &  * & 0 \\ 0 & * & 0 \end{pmatrix},\quad
    k_\tau = \begin{pmatrix} 0 & 0 & * \\ 0 & 0 &  * \\  0 & 0 &  * \end{pmatrix} \, ,
\end{align}
from which all patterns can be constructed. Here, ``$*$'' denote non-zero entries whose  parametric flavor dependence is given by the $U(1)_\text{FN}$.
Table \ref{tab:patterns_single_coupling} shows the resulting patterns for the Yukawa matrices $Y_{AL} (Y_{\bar AL})$ and $Y_{AE} (Y_{\bar AE})$ as linear combinations of the $k_\ell$-matrices, $\ell=e, \mu, \tau$.
\begin{table}[h]
    \centering
    \begin{tabular}{ccccccc}
	    \hline
	    & $k_e$ & $k_\mu$ & $k_\tau$ & $\left[ \Delta \right]_{A_4}$ & $\left[A\Delta B\right]_{Z_3}$ & name \\
	    \hline

	    \multirow{5}{*}{$Y_{AL} (Y_{\bar AL})$} & $c_\ell$ & $0$ & $0$ & $1\phantom{''}$ & \multirow{3}{*}{$0$} & $L_e$ \\
	    & $0$ & $c_\ell$ & $0$ & $1''$ & & $L_\mu$ \\
	    & $0$ & $0$ & $c_\ell$ & $1'\phantom'$ & & $L_\tau$ \\
	    \cline{2-7}
	    & $c_\nu $ & $c_\nu $ & $c_\nu$ & $1$, $1'$, $1''$ & $1$ & $L_d$ \\
	    \cline{2-7}
	    & $c_\nu \kappa$ & $c_\nu \kappa $ & $c_\nu \kappa$ & $1$, $1'$, $1''$ & $2$ & $L_{d'}$ \\
	    \hline
	    \hline
	    \multirow{7}{*}{$Y_{AE} (Y_{\bar AE})$} & $1$ & $0$ & $0$ & $1\phantom{''}$ & \multirow{3}{*}{$0$} & $R_e$ \\
	    & $0$ & $1$ & $0$ & $1''$ & & $R_\mu$ \\
	    & $0$ & $0$ & 1 & $1'\phantom'$ & & $R_\tau$ \\
	    \cline{2-7}
	    & $\kappa $ & $0$ & $\kappa^\prime$ & $1\phantom{''}$ & \multirow{3}{*}{$1$} & $R_{e\tau}$ \\
	    & $\kappa^\prime$ & $\kappa $ & $0$ & $1''$ & & $R_{e\mu}$ \\
	    & $0$ & $\kappa^\prime $ & $\kappa$ & $1'\phantom'$ & & $R_{\mu\tau}$ \\
	    \cline{2-7}
	    & $\kappa^{\prime 2} $ & $\kappa^{\prime 2} $ & $\kappa^{\prime 2}$ & $1$, $1'$, $1''$ & $2$ & $R_d$ \\
	    \hline
    \end{tabular}
    \caption{Patterns for the Yukawa matrices of  left-handed (upper part) and right-handed  (lower part) leptons for leptoquarks in singlet representations of $A_4$.
    Additional FN-factors apply and are given in Eq.~(\ref{eq:yab}).}
    \label{tab:patterns_single_coupling}
\end{table}
For instance, the $R_{e \mu}$ pattern corresponds to $\kappa^\prime k_e+\kappa k_\mu$ amended by FN-factors that depend on the leptoquark scenario
and can be taken from Eqs.~(\ref{eq:yab}) and (\ref{eq:charges})
 \begin{align} \label{eq:Rup}
R_{e \mu }(UE)=  \begin{pmatrix} \kappa^\prime  \lambda^8 & \kappa  \lambda^6 & 0 \\ \kappa^\prime  \lambda^6 & \kappa \lambda^4 & 0 \\  \kappa^\prime \lambda^4 & \kappa \lambda^2 & 0 \end{pmatrix} \, , ~
R_{e \mu }(\bar UE)= \begin{pmatrix} \kappa^\prime  \lambda^0 & \kappa  \lambda^2& 0 \\ \kappa^\prime  \lambda^2 & \kappa \lambda^0 & 0 \\  \kappa^\prime \lambda^4 & \kappa \lambda^2 & 0 \end{pmatrix} \, .
\end{align}
It is manifest from these patterns that generational hierarchies can be inverted relative to the ones of the  fermion mass terms.
Neglecting terms of order $\lambda^2$ the pattern $R_{e \mu }(\bar UE)$, which can appear in the $\tilde V_1$ scenario, 
closely resembles patterns leading to sizable LFV in
rare charm decays  \cite{deBoer:2015boa}.
Contributions with  $\left[A\Delta B\right]_{Z_3}=2$ arise at second order in the $A_4$-flavon expansion, and yield democratic patterns, $L_{d^\prime}$ and $R_d$, see Table \ref{tab:patterns_single_coupling}.

For the leptoquarks $S_1, S_2, V_1$ and $V_2$  both  left- and right-handed couplings can be  present simultaneously.
Since the lepton and quark mass terms must be $Z_3$-invariant, both interaction terms of the respective leptoquark must have identical  $Z_3$-charge
\begin{equation}
	[QL\Delta]_{Z_3} = [UE\Delta]_{Z_3},\ [\bar UL\Delta]_{Z_3} = [\bar QE\Delta]_{Z_3},\ [\bar QL\Delta]_{Z_3} = [\bar DE\Delta]_{Z_3},\ [DL\Delta]_{Z_3} = [QE\Delta]_{Z_3} 
\end{equation}
and  identical $A_4$-charge of the $\Delta$. Possible correlations can be read-off from Table \ref{tab:patterns_single_coupling}.
For instance,  $\left[A\Delta B\right]_{Z_3}=0$ gives $L_\ell$ and $R_\ell$, with the joint lepton flavor $\ell$ fixed by $\left[ \Delta \right]_{A_4}$. Note, that  there is an overall hierarchy 
between the left-handed couplings, which go with $c_\ell$, and the right-handed ones, which are order one.
Another possibility is $\left[A\Delta B\right]_{Z_3}=1$, which induces $L_d$ together with one of the $R_{\ell \ell^\prime}$ ones, where the selection of leptons is again fixed by $\left[ \Delta \right]_{A_4}$. For  $c_\nu \ll \kappa, \kappa^\prime$ the democratic and phenomenologically  dangerous pattern can be suppressed relative to the  lepton singlet couplings.

Quite generally, and  beyond the explicit   $A_4 \times Z_3$ model,  the  flavon VEV  suppression in leptoquark couplings to
lepton doublets  cannot be avoided, once the three generations of doublets  are in a triplet representation of  the non-abelian group  in order to give the Pontecorvo-Maki-Nakagawa-Sakata (PMNS)-matrix. This feature is of course manifest \cite{Varzielas:2015iva} in the $A_4 \times Z_4$ model \cite{Altarelli:2009kr,Varzielas:2012ai}.
Requiring invariance of the term $A \Delta L$ one  therefore needs an insertion of a triplet flavon VEV. The other alternative would be to make the leptoquark a triplet, which
leads to a democratic pattern and does not give rise to LNU.
Note, in see-saw models, terms with right-handed neutrinos, which are triplets of $A_4$ and carry $Z_2=2$ \cite{Altarelli:2009kr} result in VEV-suppressed, democratic patterns.

\subsubsection{Mass basis rotation} 

We consider modifications of the patterns derived in the flavor basis from changing to the mass basis.
The corresponding  transformations of the fermion fields by the unitary matrices $U,V$ read
\begin{align}
	u_\text{L} &\to V_u u_\text{L}   \, , & d_\text{L} & \to V_d d_\text{L}    \, ,   \\
	u_\text{R} &\to U_U u_\text{R} \, , & d_\text{R} &\to U_D d_\text{R} \, ,   \\
	\ell_\text{L} &\to U_L \ell_\text{L} \, , & \nu_\text{L} &\to U_\nu \nu_\text{L}  \, , \\
	\ell_\text{R} &\to U_E \ell_\text{R} \, , & & 
\end{align}
from which the Cabibbo-Kobayashi-Maskawa (CKM) and PMNS mixing matrices are obtained as
\begin{align}
	V_\text{CKM} = V_u^\dagger V_d \, ,\quad
	V_\text{PMNS} = U_L^\dagger U_\nu\, .
\end{align}
In leptoquark models also other combinations become physical. In particular,
the  leptoquark yukawas transform as
\begin{align}
	Y_{AB} \to U_{A}^T Y_{AB} U_{B}  \, , \quad 
	Y_{\bar{A}B}  \to U_{A}^\dagger Y_{\bar{A}B} U_{B} \, .
\end{align}
Quark rotations therefore only mix rows, whereas  lepton rotations only mix columns.

The parametric dependence of the rotation matrices in the quark sector can be obtained by perturbative diagonalization \cite{Nir:2002ah}  as
\footnote{For the charges given in Eq.~(\ref{eq:charges}) some tuning has to be done  to recover $V_{us}$.}
\begin{align} \nonumber
	\left( V_u \right)_{ij} \sim \left( V_d \right)_{ij} &\sim \lambda^{|q(Q_i)_-q(Q_j)|}   \, , \\ \label{eq:quarkrot}
	\left( U_U \right)_{ij} &\sim \lambda^{|q(U_i)-q(U_j)|}   \, ,  \\
	\left( U_D \right)_{ij} &\sim \lambda^{|q(D_i)-q(D_j)|} \, .     \nonumber
\end{align}
The resulting mixing of the  rows does not spoil the patterns  as the hierarchical suppression of the leptoquark yukawas stays parametrically intact.
Note, that this does not hold true anymore for quarks charged non-trivially under $A_4$, as discussed in Section \ref{sec:nontrivial}.

Since the transformations $U,V$ are unitary and neutrinos are  inclusively reconstructed  in collider experiments, the rotation $V_\nu$ has no impact on such observables.
Furthermore, in the $A_4 \times Z_3$ framework considered in this work, the charged lepton Yukawa matrix $Y_\ell$ is already diagonal at  leading order.
However, higher order flavon insertions can induce non-diagonal entries in $Y_\ell$  \cite{Altarelli:2005yx}.
We discuss this in the next section together with other higher order effects.

\subsubsection{Higher order flavon  corrections}

It is easy to compute $Y_\ell$ including next-to leading order corrections
\begin{equation}
	Y_\ell \sim c_\ell \left[ \begin{pmatrix} \lambda^4 & 0 & 0 \\ 0 & \lambda^2 & 0 \\ 0 & 0 & 1 \end{pmatrix} + \delta \begin{pmatrix} \lambda^4 & \lambda^2 & 1 \\ \lambda^4 & \lambda^2 & 1 \\ \lambda^4 & \lambda^2 & 1 \end{pmatrix}  \right] \sim c_\ell \begin{pmatrix} \lambda^4 & \delta \lambda^2 & \delta \\  \delta \lambda^4 & \lambda^2 & \delta \\  \delta \lambda^4 &  \delta \lambda^2 & 1 \end{pmatrix} \, ,
\end{equation} 
from which the rotation matrices follow as, using perturbative diagonalization \cite{Nir:2002ah},
\begin{align}
	U_L  \sim \begin{pmatrix} 1 & \delta & \delta \\ \delta & 1 & \delta \\ \delta & \delta & 1 \end{pmatrix} \, , \quad 
	U_E  \sim  \begin{pmatrix} 1 & \delta \lambda^2 & \delta \lambda^4 \\ \delta \lambda^2 & 1 & \delta \lambda^2 \\ \delta \lambda^4 & \delta \lambda^2 & 1 \end{pmatrix} \, .
\end{align} 
Here, we  introduced a parameter $\delta <1$, of the order $({\rm VEV})^2$,
\begin{equation}
\delta\sim  \text{max} \left( \frac{c_\nu^3}{c_\ell}, \frac{c_\nu \kappa^2}{c_\ell}, \frac{c_\nu \kappa \kappa^\prime}{c_\ell}, \frac{c_\nu \kappa'^2}{c_\ell} \right)  \, .
\end{equation}

The effect of transforming the left-handed charged leptons is therefore ${\cal{O}}(\delta)$, at second relative order in the flavon expansion and modifies
 $Y_{AL, \bar A L}$. This implies, for instance, for the tau-isolation
pattern
\begin{align} \label{eq:Ltau} 
L_{ \tau}(UL, \bar U L, QL, \bar QL) \to   c_\ell \begin{pmatrix}  \delta\lambda^4  & \delta\lambda^4  &  \lambda^4 \\   \delta\lambda^2 & \delta\lambda^2 &  \lambda^2\\  \delta &  \delta&  \lambda^0\end{pmatrix}
 \, . 
\end{align}

Rotations stemming from the right-handed leptons contribute at higher orders in $\lambda$.
{\it E.g., } this effect  modifies  single and double lepton isolation patterns  in $Y_{AE, \bar A E}$  such as
those given in Eq.~(\ref{eq:Rup}) 
\begin{align} 
R_{e \mu }(\bar UE) \to \begin{pmatrix} \kappa^\prime  \lambda^0 & \kappa  \lambda^2& \delta \lambda^4 (\kappa + \kappa') \\ \kappa^\prime  \lambda^2 & \kappa \lambda^0 &  \delta \lambda^2 \kappa \\  \kappa^\prime  \lambda^4 & \kappa \lambda^2 & \delta \lambda^4 \kappa \end{pmatrix} \, .  \label{eq:Remu}
\end{align}
For the $R_\tau$-pattern mass rotation effects  amount to 
\begin{align} \label{eq:Rtau}
R_{ \tau}(UE, \bar U E,QE, \bar QE) \to  \begin{pmatrix} \delta \lambda^8  & \delta \lambda^6 &  \lambda^4 \\ \delta \lambda^6 &  \delta \lambda^4 &  \lambda^2\\ \delta \lambda^4 & \delta \lambda^2 &  \lambda^0\end{pmatrix}  \, , 
 \quad 
 R_{ \tau}(DE, \bar D E) \to  \begin{pmatrix} \delta \lambda^7  & \delta \lambda^5 &  \lambda^3 \\ \delta \lambda^6 &  \delta \lambda^4 &  \lambda^2\\ \delta \lambda^6 & \delta \lambda^4 &  \lambda^2\end{pmatrix}  \, .
\end{align}

The patterns given  in Table \ref{tab:patterns_single_coupling} receive in addition  direct contributions from higher order flavon insertions.
The single lepton generation isolating patterns with  coupling to left-handed fermions, $L_\ell$, receive corrections from replacing  $\phi_{\ell}$ with $\phi_{\nu}$ plus two additional 
$A_4$-singlet flavons or two insertions of $\phi_\nu$. These  contributions are $\mathcal{O} (c_\nu^3/c_\ell) $ and $\mathcal{O} (c_\nu \kappa^{(\prime)2} /c_\ell)$, respectively,  and universal  for all entries modulo the FN-charges \cite{Varzielas:2015iva}.
In terms of  $\delta$ introduced before these higher order effects amount to the same as what we got from the mass basis rotation, Eq.~(\ref{eq:Ltau}).
The democratic pattern $L_d$ is subject to next-to leading order corrections  from  $\phi_\nu \to \phi_\ell$ plus one additional $A_4$-singlet. However, because of the unknown $\mathcal{O} (1)$ coefficients, these corrections are immaterial.

The explicit higher order flavon corrections  to  the patterns of right-handed leptons arise universally for each entry at third order:
two times $\phi_\nu$ plus one singlet flavon or three singlet flavons.
Denoting $\delta^\prime={\cal{O}}{(\rm VEV}^3)$, for $R_\tau$,
\begin{align} \label{eq:Rtauflavon}
R_{ \tau}(UE,\bar QE)  \! \to \! \!  \begin{pmatrix} \delta^\prime \lambda^8  & \delta^\prime \lambda^6 & \lambda^4 \\ \delta^\prime \lambda^6 &  \delta^\prime \lambda^4 &  \lambda^2\\ \delta^\prime \lambda^4 & \delta^\prime \lambda^2 &  \lambda^0\end{pmatrix} \! ,
R_{ \tau}(\bar U E, QE) \! \to  \!  \!  \begin{pmatrix} \delta^\prime \lambda^0 & \delta^\prime \lambda^2 & \lambda^4 \\ \delta^\prime \lambda^2 &  \delta^\prime \lambda^0 &  \lambda^2\\ \delta^\prime \lambda^4 & \delta^\prime \lambda^2 &  \lambda^0\end{pmatrix} \! ,
R_{ \tau}(\bar D E)  \! \to  \!  \! \begin{pmatrix} \delta^\prime \lambda  & \delta^\prime \lambda &  \lambda^3 \\ \delta^\prime \lambda^2 &  \delta^\prime \lambda^0  &  \lambda^2\\ \delta^\prime \lambda^2 & \delta^\prime \lambda^0 &  \lambda^2\end{pmatrix} \! .
\end{align}
If there are cancellations between the FN charges of the quarks and leptons, these corrections can be larger than the mass rotation effect Eq.~(\ref{eq:Rtau}).
For phenomenology one therefore has to take the maximum of each entry of Eq.~(\ref{eq:Rtau}) and (\ref{eq:Rtauflavon}).
Similarly, for  the double lepton isolation patterns 
\begin{align}
R_{e \mu }(\bar UE, \bar QE) \to \begin{pmatrix} \kappa^\prime  \lambda^0 & \kappa  \lambda^2& \delta^{\prime \prime} \lambda^4  \\ \kappa^\prime  \lambda^2 & \kappa \lambda^0 &  \delta^{\prime \prime} \lambda^2  \\  \kappa^\prime  \lambda^4 & \kappa \lambda^2 & \delta^{\prime \prime} \lambda^0 \end{pmatrix} \, , \label{eq:Remuflavon}
\end{align}
where $\delta^{\prime\prime}={\cal{O}}{(\rm VEV}^4)$.

\subsection{Quarks non-trivial under $A_4$ \label{sec:nontrivial}}

Single quarks in a non-trivial  $A_4 \times Z_3$-representation allows to construct  further   flavor patterns for the leptoquarks. 
In Ref.~\cite{Varzielas:2015iva} this has been discussed for
$A_4 \times Z_4$ models. Here, to formally
restore  $A_4 \times Z_3$-invariance of the SM yukawa terms of  the quarks   insertions of $\xi^\prime$ are necessary.
In order to not destroy the quark masses and mixings,
the $A_4$-VEV suppression  $\kappa^\prime \sim \lambda^m$ needs to be compensated by a corresponding change in  FN-charge.
Additionally, the $Z_3$ charge of the inserted flavon fields has to be cancelled.
The following choices  leave the SM Yukawa matrices of  the quarks  intact:
\begin{align}
	\left[ A_i \right]_{A_4} &\to 1'' ,    \quad 
	\left[ A_i \right]_{Z_3} \to 1,  \quad 
	q( A_i ) \to q( A_i ) -  m \, , 
\end{align} 
or,  with two insertions,
\begin{align}
	\left[ A_i \right]_{A_4} &\to 1',   \quad 
	\left[ A_i \right]_{Z_3} \to 2,  \quad
	q( A_i ) \to q( A_i ) - 2 m \, .
\end{align} 
Here, $A$ can be any of the quark fields $\bar{Q},U,D $ of  first or second generation,  $ i = 1,2 $.
For the third generation this leads to a suppression of third generation yukawas.

The different charges for one generation of quarks lead to a mixing of rows between patterns characterized by different $\left[ A \Delta B \right]_{Z_3}$ and $\left[ \Delta \right]_{A_4}$ and a modified hierarchy in the entries ``$*$'' of the lepton flavor isolating textures, $k_\ell$.
If the quark generations $j \neq i$ are trivially charged and couple to the pattern characterized by
\begin{align}
	\left[ A \Delta B \right]_{Z_3} &= a \,  , \quad [\Delta]_{A_4}  \, , 
\end{align}
see also Table \ref{tab:patterns_single_coupling}, then the $i$th row corresponding to the non-trivially charged quark is given by  the pattern with
\begin{align}
		\left[ A \Delta B \right]_{Z_3} &= \begin{cases} (a+1)\text{mod} \ 3\text{\quad for one insertion} \\ (a+2) \text{mod} \ 3 \text{\quad for two insertions}\end{cases},
		\end{align}
and the total $A_4$-charge $[A\Delta]_{A_4}$ of the quark and the leptoquark. Note that since the FN-charge of quarks has been changed, corresponding mass basis rotations
 Eq.~(\ref{eq:quarkrot}) do matter.

Choosing $\left[ \bar{Q}_2 \right]_{A_4} = \left[ \Delta \right]_{A_4} =1''$, that is, $a=0$ and $m=2$ gives a modification of the $\mu$-isolation pattern, as 
\begin{align}
    \tilde L_\mu (QL ) &=   \begin{pmatrix} 0 & c_\ell\lambda^4 & 0 \\ c_\nu\kappa & c_\nu\kappa & c_\nu\kappa  \\ 0 &  c_\ell\lambda^0 &0\end{pmatrix} ,  \quad
      \tilde L_\mu (\bar QL ) =   \begin{pmatrix} 0 & c_\ell\lambda^4 & 0 \\ c_\nu  & c_\nu  & c_\nu   \\ 0 &  c_\ell\lambda^0 &0\end{pmatrix} , 
\end{align}
where for  $QL$  the second row has $\left[ A \Delta B \right]_{Z_3}=2$ and correspondingly couples to $L_{d^\prime}$  and for $\bar QL$
 the second row has $\left[ A \Delta B \right]_{Z_3}=1$ and correspondingly couples to $L_{d}$.
 Including mass basis corrections
\begin{align} \label{eq:lmutilde}
  \tilde L_\mu (QL ) \!\! \to \!\! \begin{pmatrix}c_{\nu} \kappa \lambda^{2} & c_{\ell} \lambda^{4} + c_{\nu} \kappa \lambda^{2} & c_{\nu} \kappa \lambda^{2}\\ c_{\nu} \kappa & c_{\ell} \lambda^{2} + c_{\nu} \kappa & c_{\nu} \kappa\\ c_{\ell} \delta + c_{\nu} \kappa \lambda^{2} & c_{\ell} & c_{\ell} \delta + c_{\nu} \kappa \lambda^{2}\end{pmatrix} \, , 
    \tilde L_\mu (\bar QL ) \!\! \to\!\! \begin{pmatrix}c_{\nu} \lambda^2 &  c_{\nu} \lambda^2 & c_{\nu} \lambda^2 \\ c_{\nu}  & c_{\ell} \lambda^{2} + c_{\nu}  & c_{\nu} \\\ c_{\ell} \delta + c_{\nu}  \lambda^{2} & c_{\ell} & c_{\ell} \delta + c_{\nu}  \lambda^{2}\end{pmatrix} \, ,
\end{align}
where we note that due to Eq.~(\ref{eq:charges}) the FN-suppression of the  first row is $\lambda^2$. 
The FN-suppression of the second row present in $L_\mu$ is  in $\tilde L_\mu$  turned  into  a VEV-suppression.
The $\tilde L_\mu$- patterns are relevant for $b \to s \mu \mu$ processes.
Similarly, modifications of $\tau$-isolation patterns can be obtained for  $\left[ \bar{Q}_2 \right]_{A_4} =1''$, $\left[ \bar{Q}_2 \right]_{Z_3} =1$,  $\left[ \Delta \right]_{A_4} =1'$, that is, $a=0$ and $m=2$, as 
\begin{align} \label{eq:max}
    \tilde L_\tau ( \bar QL ) &=   
       \begin{pmatrix} 0 & 0 & c_\ell\lambda^4  \\ c_\nu  & c_\nu  & c_\nu   \\ 0 & 0& c_\ell\  \end{pmatrix} , 
\end{align}
which is an example for a pattern that potentially maximizes the effect from doublet quarks and leptons in $R_D,R_{D*}$.
Relevant leptoquarks are $V_1$ and $V_3$. 
After mass basis rotations 
\begin{align}
    \tilde L_\tau ( \bar QL ) & \to 
       \begin{pmatrix} \lambda^2 c_\nu  & \lambda^2 c_\nu  & \lambda^2 c_\nu   \\ c_\nu  & c_\nu  & c_\nu   \\ \lambda^2 c_\nu + \delta c_\ell  &  \lambda^2 c_\nu + \delta c_\ell  & c_\ell  \end{pmatrix}  \, .
\end{align}
 For  $V_1$ and $V_3$ constraints from $\mu-e$-conversion  data apply as
$\lambda^4 c_\nu^2 \lesssim 7 \cdot 10^{-7}(M/ {\rm TeV})^2 $  and  $\lambda^4 c_\nu^2 \lesssim 3.5  \cdot 10^{-7}(M/ {\rm TeV})^2$, respectively \cite{deBoer:2015boa},  therefore, $c_\nu \lesssim 0.02 (M/{\rm TeV})$. $M$ denotes the mass of the leptoquark.
Both $V_1$ and $V_3$ are also constrained by LFV kaon processes $ s \to d  e \mu$ 
 \cite{Carpentier:2010ue} $c_\nu^2 \lambda^2  \lesssim  5 \cdot 10^{-6} (M/ {\rm TeV})^2$, that is, $c_\nu \lesssim 0.01(M/{\rm TeV})$, somewhat stronger than $\mu-e$-conversion.
This prohibits noticeable effects in $b \to s \mu \mu$ transitions, which are induced at parametrically the same order of magnitude as the kaon decay.
Constraints  on scalar Wilson coefficients, which involve $\tilde L_\tau \cdot R_\tau$, exist from the $B_s \to \mu \mu$ branching ratio. They read $ \delta c_\nu < 2 \cdot 10^{-3} (M/{\rm TeV})^2$ and can be evaded naturally for 
$\delta \lesssim 0.1$.

A similar pattern can be obtained for $\bar U L$-couplings, by charging up-quark singlets non-trivially, however, with an additional suppression of the second row by $\kappa$
relative to Eq.~(\ref{eq:max}). $\tilde L_\tau (\bar U L)$ is relevant for model $S_2$. Including mass basis corrections,
\begin{align} \label{eq:maxU}
    \tilde L_\tau ( \bar UL ) & \to 
       \begin{pmatrix} \lambda^2 \kappa  c_\nu  & \lambda^2 \kappa c_\nu  & \lambda^2  \kappa c_\nu   \\ \kappa c_\nu  & \kappa c_\nu  & \kappa c_\nu   \\ \lambda^2 \kappa c_\nu + \delta c_\ell  &  \lambda^2 \kappa c_\nu + \delta c_\ell  & c_\ell  \end{pmatrix}  \, . 
\end{align}
There are no kaon bounds on $\tilde L_\tau (\bar U L)$.
The branching ratios of $D \to \mu \mu$ and similarly  $D \to \pi \mu \mu $ decays imply \cite{deBoer:2015boa} $\kappa^2 c_\nu^2 \lambda^2  \lesssim 0.06(M/ {\rm TeV})^2$, that is effectively no constraint, $\kappa c_\nu \lesssim 1$. 
$\mu-e$-conversion  data  \cite{deBoer:2015boa} impose $\kappa c_\nu \lesssim 0.02 (M/{\rm TeV})$.

\FloatBarrier

\section{Flavor Phenomenology  \label{sec:phen}}

The flavor patterns obtained in Section \ref{sec:A4} can  be used directly for  predictions in  flavor physics. Contributions to dimension six operators induced by  tree level leptoquark exchange
can be read-off from Tables \ref{tab:effectivevertices_scalars}   and \ref{tab:effectivevertices_vectors}     for scalar and vector leptoquarks, respectively,  updating  \cite{Buchmuller:1986zs} with signs and tensor operators. 
To discuss LNU in the $B$-system and explore possible signatures in charm
we additionally provide the Wilson coefficients for the  semileptonic transitions
$b \to c \tau \nu$ in Table \ref{tab:btoc_wilsons}, for $b \to s \ell \ell, \nu \bar \nu$ in Table \ref{tab:btos_wilsons} and for $c \to u \ell \ell, \nu \bar \nu$ in Table \ref{tab:ctou_wilsons}.
We discuss in Section \ref{sec:cc} leptoquark effects in $B \to D^{(*)} \ell \nu$ decays and in Section \ref{sec:RK}  LNU signals in $b \to s \ell \ell$ processes within flavor models. In Section \ref{sec:charm} we work out signatures for  charm and kaon decays and $\mu-e$ conversion.

\subsection{Leptoquark effects in $B \to D^{(*)} (e,\mu,\tau) \nu$  decays\label{sec:cc}}

Charged current-induced decays $B \to D^{(*)}  (e, \mu,\tau) \nu$ have reached a lot of attention due to the  anomalies in the observables $R_D$ and $R_{D^*}$
\begin{align}
	R_{D^{(*)}}= \frac{\mathcal B(B\to D^{(*)}\tau\nu_\tau)}{\mathcal B(B\to D^{(*)}\ell\nu_\ell)}   \, ,
	\label{eq:RDdef}
\end{align}
where in the denominator $\ell=\mu$ at LHCb  and $\ell=e,\mu$ at Belle  and BaBar.
In Table \ref{tab:RD} experimental findings and SM  predictions for $R_D, R_{D^*}$ and the $\tau$-polarization  $P_\tau$, as measured in the rest frame of the
 $B$-meson,
\begin{align}
P_\tau =\frac{\mathcal B^+ - \mathcal B^-}{\mathcal B^+ + \mathcal B^-} \, , 
\end{align}
 are given. Formulae for the branching ratios involving left- and right-polarized $\tau$-leptons, $\mathcal B^-$ and $\mathcal B^+$, respectively, are given in Appendix \ref{app:cc}.
 SM predictions for $P_\tau (D^*)$ and $P_\tau (D)$ are obtained by using the form factors of Refs.~\cite{Sakaki:2013bfa} and  \cite{Na:2015kha}, respectively. 
 Our SM value of $P_\tau (D^*)$ is in very good agreement with  the one quoted in \cite{Abdesselam:2016xqt}, $P_\tau (D)$ has larger uncertainties due to the lattice form factors.
   We define $\hat R_{D^{(*)}} \equiv   R_{D^{(*)}}/ R_{D^{(*)}}^{\rm SM}$, $\hat P_\tau \equiv P_\tau/ P_\tau^{\rm SM}$, and use in our  analyses
 \begin{align} \label{eq:Rave}
	 \hat R_D^{\rm exp}=1.35 \pm 0.17 \, , \quad \quad   \hat R_{D^*}^{\rm exp}=1.23 \pm  0.07\, \, , \quad \quad  \hat P_\tau(D^*)^{\rm exp}=0.75 \pm 1.09 \, .
 \end{align}
 Note, $\hat R_{D^*}^{\rm exp}=1.26 \pm  0.07$ without  \cite{Abdesselam:2016xqt}.

\begin{table}
    \centering
    \begin{tabular}{cr|c|c|c|c}
       &     & $R_D$ & $R_{D^*}$ & $P_\tau (D^*)$ & $P_\tau (D)$ \\ \hline
    BaBar & \cite{Lees:2012xj} & $0.440 \pm 0.058 \pm 0.042$ & $0.332 \pm 0.024 \pm 0.018$ & - & - \\
    Belle & \cite{Huschle:2015rga}  & $0.375 \pm 0.064 \pm 0.026$ & $0.293 \pm 0.038 \pm 0.015$ & - & -\\
     Belle     & \cite{Sato:2016svk} & - & $0.302 \pm 0.030 \pm 0.011$ & - & -\\
     Belle     &  \cite{Abdesselam:2016xqt}& - & $0.270 \pm 0.035 ^{+0.028}_{-0.025}$ & $-0.38 \pm 0.51 ^{+0.21}_{-0.16}$ & - \\
    LHCb  & \cite{Aaij:2015yra}  & - & $0.336 \pm 0.027 \pm 0.030$ & - & - \\ \hline
    average$^\dagger$ &     &  $0.406 \pm 0.050$ & $0.311 \pm 0.016$ &  & \\ \hline
    SM &   & $0.300 \pm 0.008$  \cite{Na:2015kha}  & $0.252 \pm 0.003$ \cite{Fajfer:2012vx} & $-0.497 \pm 0.011$  & $0.330 \pm 0.023$
    \end{tabular}
    \caption{Experimental results  and SM predictions for $R_D^{(*)}$ and the $\tau$-polarization. $^\dagger$Error weighted average; we added statistical and systematical uncertainties in quadrature.  For $R_{D^*}^\text{ave}$ we used
     \cite{Lees:2012xj,Huschle:2015rga,Sato:2016svk,Aaij:2015yra,Abdesselam:2016xqt}. Without   \cite{Abdesselam:2016xqt},  $R_{D^*}^\text{ave}=0.317 \pm 0.017$. 
         }
        \label{tab:RD}
\end{table}

\subsubsection{Vector-like contributions \label{sec:LL}}

We begin with some general considerations on the order of magnitude of  leptoquark effects induced by a dimension six operator with doublet quarks and leptons, 
$\mathcal{O}_{V_1}$, see Appendix \ref{app:cc} 
for details.  
Such a vector-type operator is induced for the representations $V_3,S_3$, and, together with a  scalar one, $\mathcal{O}_{S_1}$,  for  $V_1$.
Leptoquark  $S_1$ also induces  $\mathcal{O}_{V_1}$, but at the same time scalar and tensor contributions; their effects are discussed in Section \ref{sec:flip}.
Employing  the expressions in  Appendix \ref{app:cc} and taking only linear BSM effects into account, one obtains, schematically,
\begin{align} \nonumber
\hat R_{D^*} -1 & \simeq 2  {\rm Re} ( C_{V_1}^\tau -C_{V_1}^\ell) =  
2 \, n(\Delta) \,  {\rm Re}  \left(  Y Y^*|_\tau -YY^*|_\ell  \right) \frac{ \sqrt{2} }{4 G_F  V_{cb} M^2} \\
& \simeq 1.5 \, n(\Delta)  \, {\rm Re}  \left(  Y Y^*|_\tau -YY^*|_\ell  \right) \ \left(  \frac{\mbox{TeV}}{ M} \right)^2 \, . \label{eq:Rdstar}
\end{align}
Here $n(\Delta)=-1/2, +1, -1$ are Fierz factors for  $S_3,V_1,V_3$, respectively. In $\hat R_{D^*}$  contributions from $ \mathcal{O}_{S_1}$ are $O( 10 \%)$,  and the expression holds for  $V_1$ at this level. For $S_3,V_3$ holds exactly  $\hat R_D=\hat R_{D^*}$.

Confronting  Eq.~(\ref{eq:Rdstar})  to data (\ref{eq:Rave}), one  obtains
\begin{align}
{\rm Re}  \left(  Y Y^*|_\tau -YY^*|_\ell  \right)   \simeq \frac{0.2 \pm 0.05}{n(\Delta)}   \left(  \frac{M}{\mbox{TeV}}\right)^2  \, . 
\end{align}
We learn that,  model-independently,  {\it i})  $M \lesssim 3 $ TeV or perturbativity breaks down and {\it ii})  to avoid collider search limits for ``third generation leptoquarks''  decaying to 
$t \tau$
$M >685$ GeV \cite{Khachatryan:2015bsa}  the yukawa couplings need to be not too suppressed,  ${\rm Re}  \left(  Y Y^*|_\tau -YY^*|_\ell  \right)    >  0.07$.
The $V_1$ leptoquark does not couple  to $t \tau$, but rather  to $t \nu$. Corresponding mass limits are similar \cite{Freytsis:2015qca}.
For scalar leptoquarks decaying  100 \% into a muon (an electron) and a jet, the limits are $M > 1160$ GeV \cite{Aaboud:2016qeg} ($M>1755$ GeV \cite{Khachatryan:2015qda}), implying ${\rm Re}  \left(  Y Y^*|_\tau -YY^*|_\ell  \right)    >  0.2$ ($> 0.5$). Limits for vector leptoquarks are model-dependent and read
 $M > 1200 -1720$ GeV ($M > 1150 -1660$ GeV) for 100 \% decays to muon plus jet  (electron plus jet) \cite{Khachatryan:2015vaa}.

A maximal prediction from  flavor models is 
\begin{align}  \label{eq:LLmax}
Y Y^*|_\tau -YY^*|_\ell   \sim  {\rm VEV}^2  \, .
\end{align}
The suppression at second order in the flavon VEV is unavoidable in couplings to
lepton doublets which are triplets of the non-abelian discrete group, and holds beyond the  $A_4 \times Z_3$ model considered here, 
see Section \ref{sec:trivial}.
An explicit realization is given by the 
model with non-trivially charged quarks, $\tilde L_\tau(\bar Q L)$, Eq.~(\ref{eq:max}), in which the FN-symmetry suppression can be evaded and  instead the
suppression is given by  the VEVs $c_\ell c_\nu$.  The latter is bounded directly by $B \to K \nu \bar \nu$-data for  leptoquark $V_3$ as $c_\ell c_\nu \lesssim  0.02 (M/{\rm TeV})^2$,
see Appendix \ref{app:bsll} for details.
In simpler flavor models, generically, there is  both $({\rm VEV})^2$ and FN-suppression, 
\begin{align} \label{eq:LLgeneric}
\ Y Y^*|_\tau -YY^*|_\ell   \sim c_\ell^2 \lambda^2 \lesssim 10^{-3}\, ,
\end{align}
as, for instance, for the $\tau$-isolation patterns, $L_\tau$, given in Eq.~(\ref{eq:Ltau}). 

We are therefore led to conclude  that flavor models cannot explain the  few$\times 0.1 $ enhancement  in $R_{D^{(*)}}$ relative to the SM  as in present days data
with vector-type operators,  that is, within the models $S_3,V_3$. $V_1$ is discussed separately in Section \ref{sec:V1}. 
On the other hand, the possible effects can show up at the level few percent for ``maximal'' and few permille  for the generic case.
The $\tau$-polarization  for BSM in  the operator $\mathcal{O}_{V_1}$ only  is SM-like, and $\hat R_{D} =\hat R_{D^{*}}$.

\subsubsection{Chirality-flipping contributions \label{sec:flip}} 

We consider now the leptoquarks  $S_1, S_2$, which induce scalar and tensor operators, $\mathcal{O}_{S_2}$ and $\mathcal{O}_T$,  respectively.
Their Wilson coefficients are related as $C_{S_2}^{\tau\nu_\tau} =  \mp r  \, C_{T}^{\tau\nu_\tau}$,  $r=7.8$,  where the upper sign (lower sign)
corresponds to $S_1$ ($S_2$) at  renormalization  scale around $m_b$, see Appendix \ref{app:cc}  
for details.

\begin{figure}
    \centering
     \includegraphics{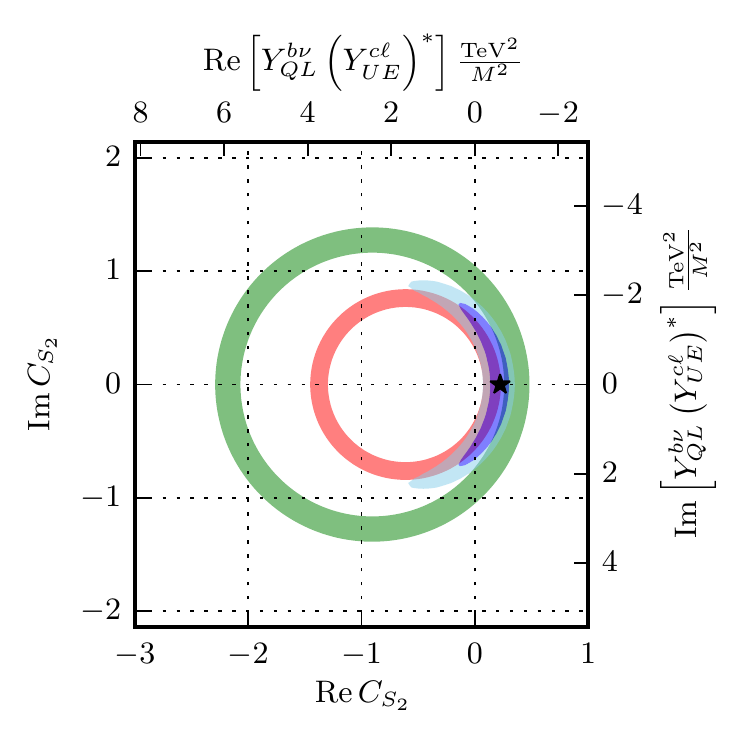}
    \includegraphics{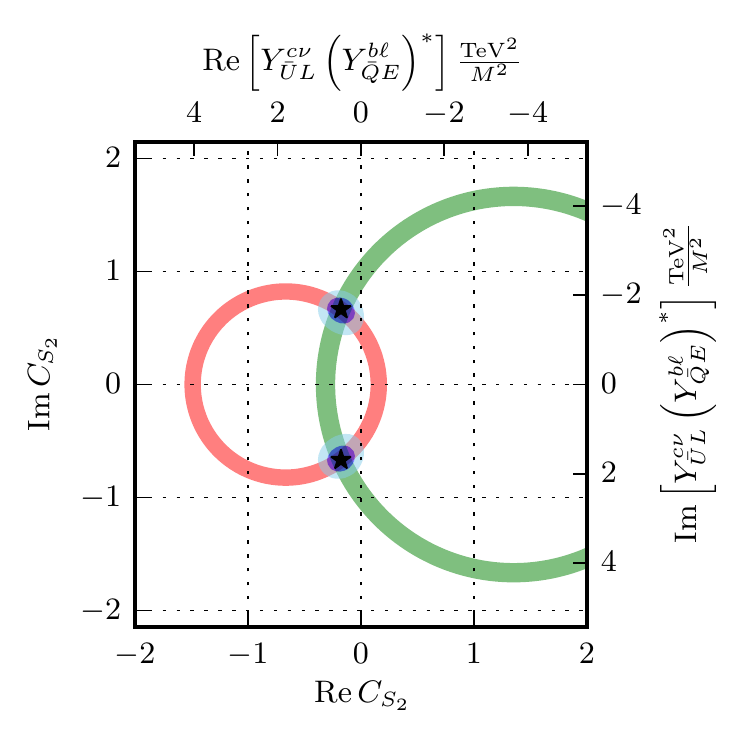}
    \caption{Preferred regions for the coupling  $Y Y^*|_\tau$ in leptoquark
      model $S_1$ (plot to the left) and   $S_2$ (plot to the right). 
   In the fit to $S_1$ we fixed   $Y_{QL} Y^*_{QL} |_\tau$ to its conservative, upper limit given by  Eq.~(\ref{eq:bnunu}).
        The red and green bands show the $1\,\sigma$ allowed regions by $R_D$ and $R_{D^*}$, respectively. Also shown is the induced Wilson coefficient $C_{S_2}^{\tau\nu_\tau}$.
        Dark and light blue bands correspond to the best fit regions at 1 and 2 $\sigma$, respectively. }
    \label{fig:LQS1_bestfit}
    \label{fig:LQ_bestfit}
\end{figure}

As  in Eq.~(\ref{eq:Rdstar}) for the  vector-type operators, we  linearize  the LNU-sensitive observables, 
\begin{align} \nonumber 
\hat R_{D^*} -1 &  \simeq - {\rm Re} ( C_{S_2}^\tau) (\hat B_{VS} ^\tau \pm \hat B^\tau_{VT}/r)  -  [ \tau \to \ell  ]  
=    {\rm Re} ( C_{S_2}^\tau)  (- 0.12 \pm 0.59)  -  [ \tau \to \ell  ]  \\
& \simeq (\mp 0.22 +  0.045)  \, {\rm Re}  \left(  Y Y^*|_\tau   \right) \ \left(  \frac{\mbox{TeV}}{ M} \right)^2 \, , \label{eq:RdstarLR} \\
 \nonumber 
\hat R_{D} -1 &  \simeq  {\rm Re} ( C_{S_2}^\tau) (\hat A_{VS} ^\tau \mp \hat A^\tau_{VT}/r)  -  [ \tau \to \ell  ]  
=    {\rm Re} ( C_{S_2}^\tau)  (1.73  \mp 0.09)  -  [ \tau \to \ell  ]  \\
& \simeq (-  0.65 \pm  0.03)  \, {\rm Re}  \left(  Y Y^*|_\tau   \right) \ \left(  \frac{\mbox{TeV}}{ M} \right)^2 \, .\label{eq:RdLR}
\end{align}
In both last rows of  Eqs.~(\ref{eq:RdstarLR}) and  (\ref{eq:RdLR}) we neglected the contributions from $\ell=e$ or $\mu$ as they enter with mass suppression relative to the $\tau$-contribution. A Fierz factor of $-1/2$ is included.
In general $\hat R_D \neq \hat R_{D^*}$ and in particular for $S_2$, corresponding to the  bottom sign,
$\hat R_D$ and $\hat R_{D^*}$ cannot be both simultaneously enhanced. 
To fit the data (\ref{eq:Rave}) in this leptoquark model, one has to go beyond the linear approximation and introduce  imaginary parts  \cite{Dorsner:2013tla,Sakaki:2013bfa}. This is illustrated in Fig.~\ref{fig:LQ_bestfit}, where we show the $1\,\sigma$ allowed regions for $R_D$ and $R_{D^*}$ for $S_1$ (plot to the left) and $S_2$ (plot to the right).
In $S_1$ also contributions to $\mathcal{O}_{V_1}$ are induced. They have not been given in Eqs.~(\ref{eq:RdstarLR}) and  (\ref{eq:RdLR}), however, to improve the fit, which is based on the full expressions, these contributions have been fixed to the conservative, upper bound on $|C_{\rm L}^{ \nu_\tau \nu_\tau}|$ allowed by the $B \to K \nu \bar \nu$ branching ratio, see Appendix \ref{app:bsll},
\begin{align} \label{eq:bnunu}
	\left|Y_{QL}^{b \nu_\tau} \left( Y_{QL}^{c \tau} \right)^*\right|  \lesssim 0.05   \left(  \frac{M}{\mbox{TeV}}\right)^2   \,   (\mbox{for}~S_1)  \, .
\end{align}
The  best fit points read
\begin{align}
Y_{QL}^{b \nu_\tau} \left( Y_{UE}^{c \tau} \right)^* &= -0.6\,    \left(  \frac{M}{\mbox{TeV}}\right)^2    \,   (\mbox{for}~S_1)  \, , \quad  \quad 
	Y_{\bar{U}L}^{c \nu_\tau} \left( Y_{\bar{Q}E}^{b \tau} \right)^* = \left(0.5 \pm 1.8  i \right)   \left(  \frac{M}{\mbox{TeV}}\right)^2  \,   (\mbox{for}~S_2) \, .
\end{align}
The  hierarchy required for $S_1$ 
\begin{align}
	\frac{ Y_{QL}^{s \nu_\tau} }{ Y_{UE}^{c \tau} } = 0.08
\end{align}
can  be explained naturally with the flavon VEVs. In $L_\tau(QL), R_\tau(UE)$ the ratio is $\sim c_\ell$.

In contrast to the lepton doublets,  the leptoquark yukawas to the lepton singlets do not require a flavon VEV insertion and
can be order one. The resulting flavor model prediction   for chirality-flipping operators is therefore subject to a single VEV suppression from the doublets only,
\begin{align} \label{eq:LRmax}
  Y Y^*|_\tau    \sim  {\rm VEV}   \, .
\end{align}
This is realized in the scalar contribution of leptoquark  $V_1$ by $\tilde L_\tau(\bar QL)$, Eq.~(\ref{eq:max}), and  the $\tau$-isolation patterns, $R_\tau(\bar D E)$, the maximum of Eq.~(\ref{eq:Rtau}) and (\ref{eq:Rtauflavon}).
The corresponding VEV  is $c_\nu$. We discuss this further in Section \ref{sec:V1}.

A maximal, pure  chirality-flipping model  is given by  leptoquark  $S_2$ with $\tilde L_\tau(\bar UL)$, Eq.~(\ref{eq:maxU}), and $R_\tau(\bar Q E)$, given by
the maximum of Eq.~(\ref{eq:Rtau}), (\ref{eq:Rtauflavon}).  This model predicts $Y Y^*|_\tau \sim \kappa c_\nu$, which is constrained by $\mu-e$-conversion  data as 
$\kappa c_\nu \lesssim 0.02 (M/{\rm TeV})$. While $Y Y^*|_\tau$ formally is of second order in the VEVs, in practice this has no effect on our analysis as we constrain $\kappa c_\nu$ experimentally  rather than employing model-specific values.
Kaon bounds are not effective in $S_2$ since only lepton singlets couple to the down quarks and the  first and second generation  block of $R_\tau$  is highly FN-suppressed.
For the former reason $b \to s \ell \ell$ processes are SM-like.

The leptoquarks $S_1$ and $S_2$ could in principle be responsible  for the  magnetic moment of the muon, as
 $\tilde L_\tau \cdot R_\tau$  patterns give rise to  chirally enhanced contributions by the top mass in the loop. 
 However, saturating  $\Delta a_\mu \sim (2-3) \cdot 10^{-9}$  \cite{Agashe:2014kda} requires  
yukawa contributions of  few permille  for $M \gtrsim 1$ TeV \cite{Davidson:1993qk,deBoer:2015boa}, while corresponding flavor model predictions are much smaller,
$c_\nu \delta \lambda^4 ,  c_\nu  \kappa \delta \lambda^4 \lesssim 4 \cdot 10^{-6} \delta \,  (M/{\rm TeV})$, respectively.

Generic predictions for chirality-flipping contributions  in flavor models are given by
\begin{align}
    Y Y^*|_\tau    \sim c_\ell \lambda^2 \lesssim 10^{-2}\, ,
\end{align}
for instance, with the patterns $L_\tau$ and $R_\tau$,
given in  Eq.~(\ref{eq:Ltau})  and  the maximum of Eq.~(\ref{eq:Rtau}), (\ref{eq:Rtauflavon}), respectively.
For leptoquark $V_2$ the contributions are induced by $L_\tau(DL)$ (or $\tilde L_\tau(DL)$) and $R_\tau(QE)$ and of the order $ Y Y^*|_\tau    \sim c_\ell \lambda^4$, 
further FN-suppressed than the generic case.

Maximal effects in $\hat R_D$ and $\hat R_{D^*}$  from chirality-flipping operators are therefore possible  at the level of a few percent ($D^*$) and reaching 0.1 ($D$) (for $S_2$), and one order of magnitude lower for the generic case. In $S_2$ an enhanced $\hat R_D$ implies a suppressed $\hat R_{D^*}$ and vice versa.

For the $\tau$-polarization, we find
\begin{equation}
	\begin{aligned}
		\hat P_\tau(D^*)-1 &\simeq -\mathrm{Re}(C_{S_2}^\tau)\left[(\hat B_{VS}^+ - \hat B_{VS}^- - \hat B_{VS}^\tau) \pm (\hat B_{V_1T}^+ - \hat B_{V_1T}^- - \hat B_{V_1T}^\tau)/r\right] \\
		&\simeq -\mathrm{Re}(C_{S_2}^\tau)(-0.36 \pm 0.19) \\
		&\simeq (0.13 \mp 0.07)\mathrm{Re}\left(YY^*\middle|_\tau\right)\left(\frac{\si{TeV}}{M}\right)^2\,,
	\end{aligned}
\end{equation}
\begin{equation}
	\begin{aligned}
		\hat P_\tau(D)-1 &\simeq \mathrm{Re}(C_{S_2}^\tau)\left[(\hat A_{VS}^+ - \hat A_{VS}^- - \hat A_{VS}^\tau) \mp (\hat A_{VT}^+ - \hat A_{VT}^- - \hat A_{VT}^\tau)/r\right] \\
		&\simeq \mathrm{Re}(C_{S_2}^\tau)(3.50 \pm 0.18) \\
		&\simeq (1.30 \pm 0.07)\mathrm{Re}\left(YY^*\middle|_\tau\right)\left(\frac{\si{TeV}}{M}\right)^2\, , 
	\end{aligned}
\end{equation}
where the upper (lower) sign corresponds to leptoquark $S_1$ ($S_2$).

\subsubsection{Leptoquark $V_1$ \label{sec:V1}}

For   $V_1$
 with   $\tilde L_\tau(\bar QL)$ and $R_\tau(\bar D E)$ \footnote{To maximize the impact on $R_{D^{(*)}}$ we allow here for FN-charges as in multi-Higgs models  such that
 $R_\tau(\bar D E)_{33} \sim \lambda^0$.}
 exist both vector-like and chirality-flipping operators
\begin{align} \nonumber 
\hat R_{D^*} -1 &  \simeq  2  {\rm Re} ( C_{V_1}^\tau) +{\rm Re} ( C_{S_1}^\tau) \hat B_{VS} ^\tau -  [ \tau \to \ell  ]  
\simeq  1.5  c_\nu   \left(   c_\ell  - 0.12   \right)      \left(  \frac{\mbox{TeV}}{ M} \right)^2  \\
& \lesssim  0.02   \left(  c_\ell  -0.12   \right)      \left(  \frac{\mbox{TeV}}{ M} \right)  \, , \\
 \nonumber 
\hat R_{D} -1 &  \simeq  2  {\rm Re} ( C_{V_1}^\tau) +{\rm Re} ( C_{S_1}^\tau) \hat A_{VS} ^\tau  -  [ \tau \to \ell  ]  
\simeq  1.5   c_\nu  \left(   c_\ell  - 1.73 \right)      \left(  \frac{\mbox{TeV}}{ M} \right)^2  \\
& \lesssim  0.03   \left(  \frac{\mbox{TeV}}{ M} \right)  \, . 
\end{align}
If the chirality-flipping contribution dominates,  both $\hat R_D$ and $\hat R_{D^*}$ can be enhanced, and at the same time differ as the
deviation from the SM is larger in $\hat R_D$.
Kaon decay constrains $c_\nu \lesssim 0.01(M/{\rm TeV})$, which has been taken into account above. Corresponding $\mu-e$ conversion bounds  are very close,
$c_\nu \lesssim 0.02(M/{\rm TeV})$.
It would therefore require the tuning of both the first and  the second quark generation coefficients to ease these constraints.
While $B \to K \nu \nu$ constraints do not apply to $V_1$ at the matching scale $\mu \sim M$,  a contribution is induced
by renormaliztion group running from $M$ to the weak scale \cite{Feruglio:2016gvd}.  Corresponding constraints are, however, weaker than the ones from kaon decays and $\mu-e$ conversion.

For the $\tau$-polarization, we find
\begin{align}
	\begin{split}
		\hat P_\tau(D^*)-1 &\simeq \mathrm{Re}(C_{S_1}^\tau)(\hat B_{VS}^+ - \hat B_{VS}^- - \hat B_{VS}^\tau) 
		 \\
		& \simeq -0.36~\mathrm{Re}(C_{S_1}^\tau) \lesssim 0.005\left(\frac{\si{TeV}}{M}\right)\,,
	\end{split} \\
	\begin{split}
		\hat P_\tau(D)-1 &\simeq \mathrm{Re}(C_{S_1}^\tau)(\hat A_{VS}^+ - \hat A_{VS}^- - \hat A_{VS}^\tau) 
	\\	&
	\simeq 3.50~\mathrm{Re}(C_{S_1}^\tau) \lesssim 0.05\left(\frac{\si{TeV}}{M}\right) \, , 
	\end{split}
\end{align}
where in the last steps we imposed kaon constraints.

\subsubsection{Synopsis of  leptoquark models for $R_{D^{(*)}}$ and the $\tau$-polarization  \label{sec:synopsis}}

\begin{figure}[h]
	\begin{center}
		\includegraphics{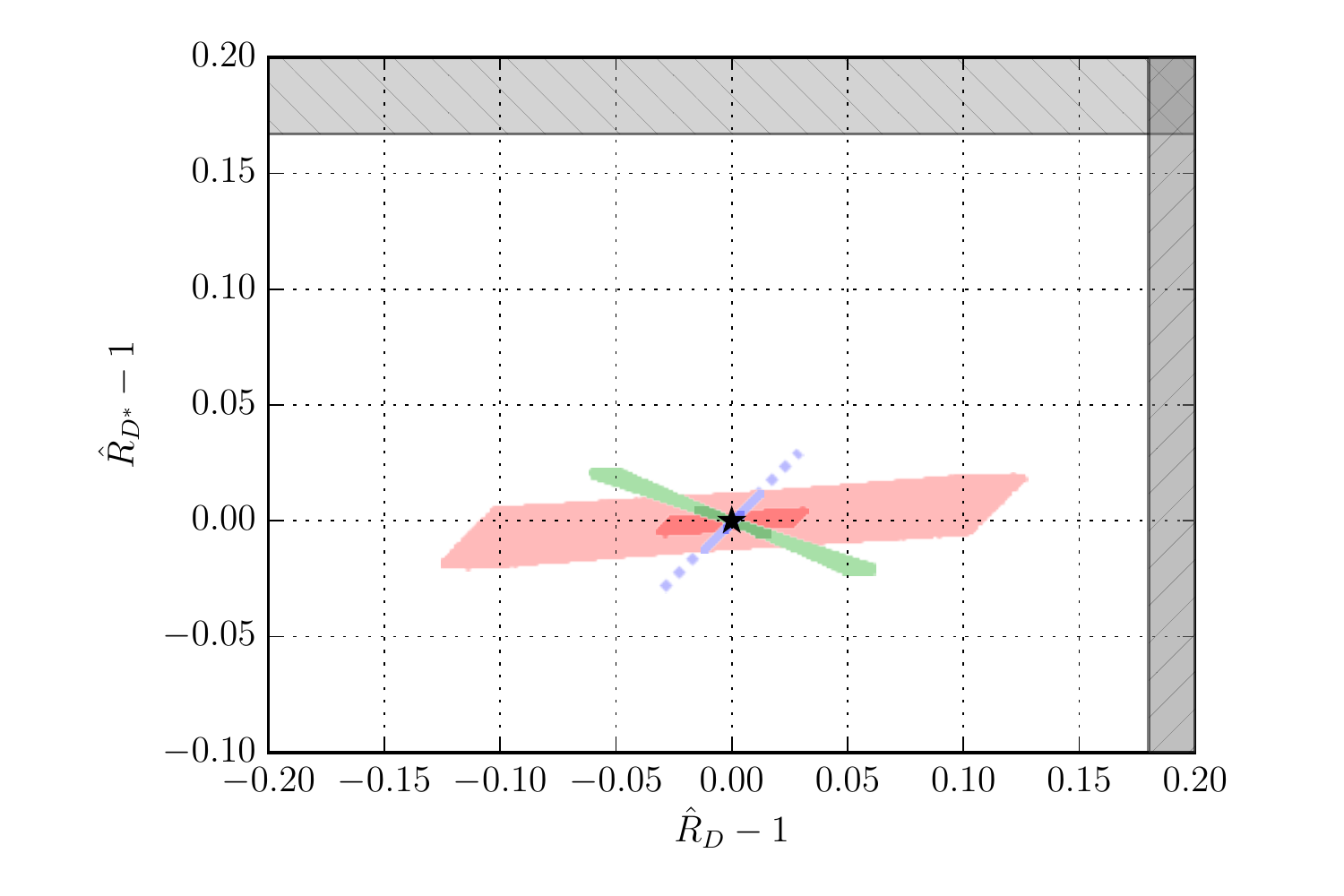}
	\end{center}
	\caption{Maximal reach of  leptoquarks  $V_1$ (red),  $V_3$ (blue)  and $S_2$ (green) in   $\hat R_{D^*}-1$  versus $\hat R_D-1$  in flavor models. 
	Darker and lighter shaded areas correspond to FN-coefficients of  $\pm 1 $ and within $\pm (1/\sqrt{2}; \sqrt{2})$, respectively. The SM is denoted by the black star. Experimental $1 \sigma$ regions (\ref{eq:Rave}) (grey) are shown only in the axes' ranges displayed.}
	\label{fig:RDvsRDstar}
\end{figure}

Maximal predictions for $\hat R_{D^{(*)}}-1$ from leptoquarks $V_1,V_3$ and $S_2$ in  flavor models are shown in Fig.~\ref{fig:RDvsRDstar}. 
Not shown are predictions for $S_{1,3}$, which  are further suppressed as they either involve three powers of flavon VEVs or FN-suppression,  as given by (\ref{eq:LLgeneric}).
The chirality-flipping contribution in $S_1$ is  constrained by kaon decays, whereas, effectively,  $S_2$ is not.
The maximal predictions are obtained with single quarks being charged non-trivially under the non-abelian  flavor symmetry.

For each leptoquark, we show two ranges, one in which the $O(1)$ coefficients from the FN-mechanism have modulus 1 (darker shaded regions), and
another one in which we allow for factors of $\sqrt{2}$ enhancement and  suppression (lighter shaded regions). The latter results effectively in enlarging  $Y Y^*$
by a factor 4, where a factor of 2 comes in directly and another one   because the low energy constraints can be eased.
For $V_3$, shown in blue, we  impose  the kaon constraint on $c_\nu$ and require $c_\ell \lesssim 0.2$. 
For $S_2$ (green) we employ the $\mu-e$-conversion bounds on $\kappa c_\nu$.
For $V_1$, shown in red,  we employ the kaon bounds on $c_\nu$ and require $c_\ell \lesssim 0.2$.
For $V_3$ we  also illustrate in dashing  the region 
which would become accessible  additionally if only the direct  bound on 
$ c_\ell c_\nu$ from  $B \to K \nu \nu$ would be used.  It shows that one is still $3 \sigma$ away from the experimental $\hat R_{D^*}$-band.

We learn that present data on $\hat R_D$ and $\hat R_{D^*}$ cannot be explained within $1.6 \sigma$  and $3 .1\sigma$, respectively. 
Difficulties in explaining  sizable BSM in  $R_{D^*}$ have also been encountered within the context of   Two-Higgs doublet models once conditions on the flavor structure are imposed \cite{Jung:2010ik}.

The Belle II  projection for the uncertainty on $R_D$ is $ 5.6 \% \, (3.4 \%)$ with $5 \, {\rm ab}^{-1} \, (50 \, {\rm ab}^{-1})$ and for
$R_{D^*}$  is  $3.2 \% \, (2.1 \%)$ for  $5 \, {\rm ab}^{-1} \, (50 \, {\rm ab}^{-1})$ \cite{Inguglia:2016acz}. This suffices to probe  all leptoquark models
on the basis of branching ratio measurements even close to SM values.

The predictions for the $\tau$-polarization are similar to the ones for $R_{D^{(*)}}$ with contributions from vector-like operators removed.
$P_\tau(D^*)$ can differ from the SM by at most a percent. Deviations from the SM in  $P_\tau(D)$ can reach up to several percent.
Present data on  the $\tau$-polarization, given in (\ref{eq:Rave}), are in agreement with the SM and are not sensitive to leptoquark flavor models yet.

\subsection{Leptoquark effects in $b \to s  \ell \ell$} \label{sec:RK}

We analyze tree level leptoquark effects  in $b \to s  \ell \ell$ within the  representations 
 $S_3, V_{1,2,3}$ and $\tilde S_2$. We do not consider $S_2$ and $\tilde S_1$ because
they induce only contributions onto operators $\bar s \gamma_\mu b \ell \gamma^\mu (1 + \gamma_5) \ell$, whose impact on  $B \to K^{(*)} \ell \ell$ branching ratios
is very small. 
We focus on explaining the measurement of $R_K$ \cite{Hiller:2003js}  by LHCb
for dilepton masses squared between 1 and 6 $\mbox{GeV}^2$ \cite{Aaij:2014ora}
\begin{align}  \label{eq:RKLHCb}
R_K |_{[1,6]}=  \frac{\mathcal B(B\to K  \mu \mu )}{\mathcal B(B\to K ee)} =0.745 \pm^{0.090}_{0.074} \pm 0.036 \, .
\end{align}
A model-independent analysis points, at $1 \sigma$, to modifications to the vector-type operators $O_{9,10}^{(\prime) \ell}$ with couplings to $\ell=e,\mu$
 as \cite{Hiller:2014yaa}, 
\begin{equation}
	0.7 \lesssim -\mathrm{Re}\left[C_9^{\text{NP}\mu}-C_{10}^{\text{NP}\mu}+ C_9^{\prime\mu}-C_{10}^{\prime\mu}  -\left( \mu \to e \right)\right] \lesssim 1.5 \, ,
	\label{eq:RK_constraint}
\end{equation}
where the operators are defined in Appendix \ref{app:bsll}.
Eq.~(\ref{eq:RK_constraint}) can be satisfied with $C_9^{\text{NP}\mu} = -C_{10}^{\text{NP}\mu} \sim Y^{b\mu}_{QL}\left(Y^{s\mu}_{QL}\right)^*$ or $Y^{s\mu}_{\bar QL}\left(Y^{b\mu}_{\bar QL}\right)^*$ with the leptoquarks $S_3$, or $V_{1,3}$, respectively,
and
\begin{equation} \label{eq:target}
	Y^{b\mu}_{QL}\left(Y^{s\mu}_{QL}\right)^* \text{ or } Y^{s\mu}_{\bar QL}\left(Y^{b\mu}_{\bar QL}\right)^* \simeq   [0.001 - 0.002]    \left(   \frac{M}{\rm TeV} \right)^2 \, . 
\end{equation}
Simple flavor patterns such as the $\mu$-isolation one $L_\mu$  \cite{Varzielas:2015iva} can accommodate this
\begin{equation}  \label{eq:A}
	Y^{b\mu}_{QL}\left(Y^{s\mu}_{QL}\right)^* \text{ or } Y^{s\mu}_{\bar QL}\left(Y^{b\mu}_{\bar QL}\right)^* \sim c_\ell^2 \lambda^2 \, ,
\end{equation}
where $c_\ell \sim  0.2 (M/\mbox{TeV})$. Considering a natural value for the VEV this points to leptoquark masses below a few TeV.
This is a stronger  bound on $M$ than the one obtained in \cite{Hiller:2014yaa} by using $B_s-\bar B_s$-mixing.
$K \to \mu \mu$  decays, induced at order $c_\ell^2 \lambda^6$,  and 
$\mu - e$-conversion, after including mass basis corrections, arising at $O(c_\ell^2 \delta \lambda^8)$, are  both  below their current limits.

With the  second quark generation transforming non-trivially under $A_4$  the FN-suppression can be evaded.
The corresponding patterns $\tilde L_\mu(QL,\bar QL)$ are given in Eq.~(\ref{eq:lmutilde}) and yield
\begin{equation} \label{eq:B}
	Y^{b\mu}_{QL}\left(Y^{s\mu}_{QL}\right)^*  \sim  c_\ell c_\nu \kappa\, , \quad  Y^{s\mu}_{\bar QL}\left(Y^{b\mu}_{\bar QL}\right)^* \sim c_\ell c_\nu \, ,
\end{equation}
for $S_3$ and $V_{1,3}$, respectively. Eq.~(\ref{eq:target}) can be accommodated with  $c_\ell \sim  0.2 (M/\mbox{TeV})$ and 
 $c_\nu \kappa \sim  0.01 (M/\mbox{TeV})$  ($S_3$) and   $c_\nu \sim  0.01 (M/\mbox{TeV})$ ($V_{1,3}$). The values of $\kappa c_\nu$ and $c_\nu$ are  set to the upper limit allowed by kaon decays, induced at order  $c_\nu^2  \kappa^2 \lambda^2$ and  $c_\nu^2 \lambda^2$, respectively. 
 As  $c_\ell$ cannot be much larger a value of  $R_K$ around  (\ref{eq:RKLHCb}) implies that the next round of LFV kaon and $\mu-e$-experiments should see a signal.

BSM effects as in  Eq.~(\ref{eq:RK_constraint}) can therefore be accommodated naturally with  $S_3,V_3$  with both $L_\mu $ and $\tilde L_\mu$-patterns.
Both leptoquark  models induce also LFV in charm, however, due to the constraints from the kaon sector, effects in charm are very  small.
In $V_1$ both left- and right-handed couplings are present.  The latter, $R_\mu(\bar D E)$, moreover  exhibits inverted flavor hierarchies, such that
kaon decays are induced at order $\lambda$, which seem to rule out $V_1$ with $\mu$-isolation patterns. However, as discussed in Section  \ref{sec:patterns},
it is viable to flip the sign of the charges $q(E)$. In this case the hierarchies in $R_\mu(\bar D E)$ would increase. Contributions to kaon decays arise at $O(\lambda^9)$, which
can be safely neglected. 
One-loop contributions to $\mu \to e \gamma$  arise in $V_1$ from  $L_\mu \cdot R_\mu$  and  $\tilde L_\mu \cdot R_\mu$, which are enhanced by the top mass. 
Corresponding constraints from ${\cal{B}}(\mu \to e \gamma) < 5.7 \cdot 10^{-13}$  \cite{Agashe:2014kda} read
$c_\ell \delta \lambda^4 , c_\nu \lambda^6 \lesssim 4 \cdot 10^{-4} (M/\mbox{TeV})^4$ \cite{Davidson:1993qk}, which are always satisfied in our  flavor models.
Therefore, after adjusting lepton singlet charges, $V_1$ provides another viable scenario for explaining sizable $R_K$.

One may employ the $\tau$--isolation patterns of model  $V_3$ discussed in the context of $R_{D^{(*)}}$  in Section \ref{sec:LL} to predict $b \to s \mu \mu$ processes.
The resulting effects are very small,  further VEV-suppressed for $L_\tau(\bar Q L)$ as $\sim \delta^2 c_\ell^2 \lambda^2$ or constrained by $b \to s \nu \nu$ and
low energy physics in $\tilde L_\tau(\bar Q L)$ as $\sim \delta c_\ell c_\nu + c_\nu^2 \lambda^2$. In either case, the effects are by orders of magnitude too small to
match Eq.~(\ref{eq:RK_constraint}).

We consider now leptoquarks  $\tilde S_2$ and $V_2$, which induce right-handed currents $C_9^{\prime\mu} = -C_{10}^{\prime\mu}$ in $b \to s \ell \ell$ transitions.
This  is disfavored by global fits to data (excluding $R_K$) on $b \to s$ transitions, which suggests predominantly  BSM in SM-type operators \cite{bobethMoriond}.
Let us nevertheless entertain this possibility as this line of research
has not reached final conclusions yet.
Right-handed currents would be signaled by $R_K\neq R_{K^*}$
\cite{Hiller:2014ula}, where $R_{K^*}$ denotes the ratio of branching fractions of $B \to K^* \mu \mu$ over  the one into electrons.
This part of our work is sensitive to the FN-charges of the down quark singlets, $q(D)$. Let us therefore be here more general than the benchmark
Eq.~(\ref{eq:charges}) and introduce $q_i \equiv q(D_i)$.
Within  the $L_\mu$-pattern, where first (second) choice corresponds to $V_2$ ($\tilde S_2$),
\begin{align}
	 Y^{b\mu}_{DL}\left(Y^{s\mu}_{D L}\right)^* & \text{ or } Y^{s\mu}_{\bar DL}\left(Y^{b\mu}_{\bar DL}\right)^* \sim c_\ell^2 \lambda^{q_3+q_2}  \simeq  [0.001 - 0.002]  ~ \mbox{or}   ~[0.002 - 0.004]  \left(   \frac{M}{\rm TeV} \right)^2 \,  ,\\
	 Y^{s\mu}_{DL}\left(Y^{d\mu}_{D L}\right)^* &  \text{ or } Y^{s\mu}_{\bar DL}\left(Y^{d\mu}_{\bar DL}\right)^* \sim c_\ell^2 \lambda^{q_2+q_1}   \lesssim   1.3 \cdot 10^{-4}  ~ \mbox{or}   ~ 2.6 \cdot 10^{-4}   \left(   \frac{M}{\rm TeV} \right)^2   \, .
\label{eq:D}
\end{align}
Explaining $R_K$ (first row) while obeying limits from $K \to \mu \mu$ \cite{Carpentier:2010ue}  (second row) strongly constrains the allowed values for the $q_i$:
\begin{align} \label{eq:q13}
\lambda^{q_1-q_3} \lesssim 0.13  \, ,
\end{align}
which prefers $q_1 \geq q_3+2$.
By perturbativity and lower  limits  on $M$, $q_2+q_3=0,1,2,3$. This is violated by the benchmark $q(D)=(3,2,2)$, which 
requires $c_\ell \sim [0.8 - 1.2] (M/\mbox{TeV})$ ($V_2$) and $c_\ell \sim  [1.2-1.7] (M/\mbox{TeV})$ ($\tilde S_2$). Both are not compatible with the flavor symmetry and
mass bounds.

In supersymmetric or multi-Higgs extensions, a viable set reads  $q(D)=(q_3+1,q_3,q_3)$, where $q_3=0,1,2,3$, all of which are in mild conflict with Eq.~(\ref{eq:q13}).
When the charges of the quark doublets and up-type quarks are also changed, two viable solutions are $q(Q)=q(U)=(3,2,0)$ and $q(D) = (2,0,0)$ or $q(D) = (3,1,1)$ \cite{Chankowski:2005qp}.
Smaller charges generically give smaller VEVs.
Choosing $q(D) = (3,1,1)$ leads to
\begin{equation} 
	Y^{b\mu}_{DL}\left(Y^{s\mu}_{D L}\right)^* \text{ or } Y^{s\mu}_{\bar DL}\left(Y^{b\mu}_{\bar DL}\right)^* \sim c_\ell^2 \lambda^2 \, ,
\end{equation}
the same FN-hierarchy as for $S_3, V_{1,3}$ obtained  in Eq.~(\ref{eq:A}). Therefore, 
$c_\ell \sim 0.2 (M/\mbox{TeV})$ ($V_2$) and  $c_\ell \sim  [0.2-0.3] (M/\mbox{TeV})$ ($\tilde S_2$), and,  consequently, leptoquark masses  should be within the few TeV-range.
$\mu-e$-conversion $\sim \delta c_\ell^2 \lambda^6$ is below experimental limits.
$\tilde S_2$ does not induce charm FCNCs at tree level. 

Similar to the situation for $V_1$ discussed previously,  in $V_2$ rapid kaon decays arise through  $R_\mu (QE)$. This can be avoided once the sign of $q(E)$ is flipped.
In this case the constraint from $\mu \to e \gamma$ reads $c_\ell \delta \lambda^3  \lesssim 10^{-4} (M/\mbox{TeV})^4$ \cite{Davidson:1993qk}, which is  always satisfied for
 perturbative $\delta.$

We learn that improved bounds on kaon decays together with $b \to s \mu \mu$ data can strongly constrain or rule out  BSM models with flavor patterns.
If solutions with down quark singlets can be ruled out, this leads to testable predictions,  the equality of LNU ratios $R_K$ and $R_{K^*}$, as
well as those of other $b\to s$ induced decay modes \cite{Hiller:2014ula}.
We  checked that the impact of  leptoquark models  explaining $R_K$ at tree level on  the observable 
${\cal{B}}(B \to D^{(*)} \mu \nu)/{\cal{B}}(B \to D^{(*)} e  \nu)$ \cite{Becirevic:2016oho}
is at permille level. We further recall that $R_K$-explaining leptoquarks can induce percent-level contributions to $b \to s \gamma$ and subsequently $b \to s \ell \ell$ spectra \cite{Hiller:2014yaa},
which can be accessed at a future high luminosity facility (with 75$\mbox{ab}^{-1}$) \cite{Hitlin:2008gf}.

LFV in $b \to s \ell \ell^\prime$ transitions  related to $R_K$   \cite{Glashow:2014iga,Varzielas:2015iva,Sahoo:2015pzk,Becirevic:2016oho} arises in the patterns 
studied in Eqs.~(\ref{eq:A}) -(\ref{eq:D}). Relative to $b \to s \mu \mu$ the effects on the amplitudes read
\begin{align}
b \to s \mu \mu &  ~ :   ~ b \to s \mu \,  (e, \tau)  ~ :  ~ b \to s  e \tau \quad \mbox{as}  \quad   1   ~ :  ~  \delta ~  :  ~ \delta^2  \quad  \quad  (L_\mu) \, , \\
b \to s \mu \mu  & ~ :   ~ b \to s \mu \,  (e, \tau)  ~ :  ~ b \to s  e \tau \quad \mbox{as}  \quad   1   ~ :  ~  1 ~  :  ~ 1  \quad \quad  ~ (\tilde L_\mu) \, .
\end{align}
The $\tilde L_\mu$ pattern predicts sizable LFV rates for leptonic and semileptonic $B_{(s)}$-decays which can be searched for  at future hadron colliders and $e^+ e^-$-machines, see
\cite{Varzielas:2015iva} for details,
\begin{align}
{\cal{B}}(B \to K  \mu^\pm e^\mp) & \sim 3 \cdot 10^{-8} \left( \frac{1-R_K}{0.23} \right)^2  , ~
{\cal{B}}(B \to K  (e^\pm, \mu^\pm) \tau^\mp)  \sim  2 \cdot 10^{-8} \left( \frac{1-R_K}{0.23} \right)^2  \, ,  \\
\frac{{\cal{B}}(B_s \to \mu^+  e^-)}{{\cal{B}}(B_s \to \mu^+  \mu^-)_{\rm SM}} & \sim   0.01 \left( \frac{1-R_K}{0.23} \right)^2  \, , \quad
\frac{{\cal{B}}(B_s \to \tau^+  (e^-, \mu^-))}{{\cal{B}}(B_s \to \mu^+  \mu^-)_{\rm SM}}  \sim  4 \left( \frac{1-R_K}{0.23} \right)^2  \, .
\end{align}

\subsection{Leptoquark effects in  rare charm and kaon decays  and $\mu-e$ conversion \label{sec:charm}}

We investigate the  implications of the  flavor patterns studied in the previous sections for  rare charm decays, $K$ decays and $\mu-e$-conversion.

Using  \cite{deBoer:2015boa} we find the following maximal upper limits, where the corresponding  scenario and pattern is indicated in parentheses:
$\mathcal{B} (D \to \pi \nu \nu ) \lesssim \num{3e-10}$  ($(S_3,V_3)$, $ L_\mu$), 
$\mathcal{B} (D \to \pi e \mu )\lesssim \num{3 e-13}$    ($V_3$, $ L_\mu$), 
$\mathcal{B} (D \to  e \mu)\lesssim \num{5 e-15}$    ($V_3$, $ \tilde L_\mu$)
and
$\mathcal{B} (D \to e \tau)\lesssim \num{7 e-17}$    ($S_2$, $ \tilde L_\tau$). $\mathcal{B} (D \to \mu \mu)$ and  $\mathcal{B} (D \to \pi \mu \mu)$ are SM-like.
Note, $\mathcal{B}^\text{SM} (D \to \mu \mu) \sim 10^{-13}$ and $\mathcal{B}^\text{SM} (D \to \pi \mu \mu) \sim 10^{-12}$ (non-resonant).
These BSM effects in charm are below present experimental  limits by many orders of magnitude. The reason is the presence of the kaon constraints, which are unavoidable once doublet quarks are involved. 
These are, however, not the largest possible signatures in charm accociated with leptoquarks in flavor models, but the largest associated with models addressing $R_{D^{(*)}}$ and $R_K$.

In scenario $\tilde V_1$ with the skewed pattern $R_{\mu e}(\bar U E)$, which could have a large impact on rare charm decays, the FN-suppression of the $(1,2)$ element is not strong enough to effectively evade the
$\mu-e$ conversion constraint, while, at the same time, keep the diagonal ones sizable.
Similarly, effects in rare charm processes from  $S_1$ with  $R_\tau (UE)$ are  $Y^{c \ell}_{ U E} Y_{ U E} ^{u \ell *}\sim \delta^{(\prime)} \lambda^{10}$, and negligible  compared to the foreseeable  experimental  sensitivity. 
With the skewed pattern $R_{e\tau}$ and leptoquark $\tilde V_1$ $\mu-e$ conversion constraints can be evaded and $c \to u e \tau$ transitions can be induced at order $\kappa \kappa' \lambda^2 \lesssim \num{1e-3}$. This leads to $\mathcal{B} (D \to e \tau) \lesssim \num{1e-13}$.

The leptoquarks $S_3$ and $V_3$ are constrained by LFV kaon decays, therefore large contributions near the experimental bound $\mathcal{B} ( K_L \to e \mu ) < \num{4.7 e-12}$ \cite{Ambrose:1998us} are expected.
The model ($S_2$, $(\tilde{L}_\tau,R_\tau)$) is bound by $\mu-e$ conversion and contributes less to rare kaon decays. 
We find $\mathcal{B} ( K_L \to e \mu ) \lesssim \num{4e-19}$.

Future $\mu-e$-conversion experiments  such as COMET \cite{Cui:2009zz} and Mu2e \cite{Bartoszek:2014mya} with  sensitivity below  $10^{-16}$, that is   2-3 orders of magnitude better than the existing  bounds,  are sensitive to the  leptoquarks $S_3$ and $V_3$ with $\tilde{L}_\mu$
discussed here. We find $\frac{\sigma ( \mu^- \text{Au} \to e^- \text{Au} )}{\sigma ( \mu^- \text{Au} \to \text{capture} )} \lesssim \num{2e-13} (\num{5e-14})$ for $V_3$  (for $S_3$).

\section{Conclusions \label{sec:con}}

We obtain  patterns for leptoquark couplings to SM fermions based on  flavor symmetries. In addition to those for lepton doublets \cite{Varzielas:2015iva}, 
we find lepton isolation patterns  for charged lepton singlets. 
These  are particularly relevant for contributions to $R_{D^{(*)}}$ involving both chiralities.
We argue on general terms that  chirality-flipping contributions are generically larger than the ones based on SM-like operators involving  doublet lepton couplings.

The flavor symmetry puts strong  constraints on the  leptoquark reach in flavor observables.
We find that  it is not possible to explain
the present data on $R_D$ and $R_D^{*}$  from tree level leptoquark exchange. The reason is  that these BSM effects  of few$\times 0.1$ are too large given  lower mass bounds on the leptoquarks, perturbativity of the flavor symmetry breaking and  flavor constraints,  importantly, 
$b \to s \nu \bar \nu$, rare kaon decays and $\mu-e$-conversion. 
We give predictions for $R_D$,  $R_{D^*}$ and the $\tau$-polarization, which are summarized in Section \ref{sec:synopsis}.
 At least the maximal leptoquark models, shown in Fig.~\ref{fig:RDvsRDstar}, can be tested at Belle II with $50 {\rm ab}^{-1}$ \cite{Aushev:2010bq}.

On the other hand, $R_K$ together with the preferred global fit in $b \to s$ observables can be
 explained naturally using muon isolation patterns  and $S_3, V_{3}$.  If one abandons the  constraints from the global fit, which prefers predominantly $V-A$-structure,
model  $\tilde S_2$ accommodates as well a few$\times 0.1$ BSM effect in semileptonic $b \to s \mu \mu$ processes.  $\tilde S_2$ also predicts $R_K \neq R_{K^*}$.

In our analysis we 
require the non-abelian flavon VEVs to remain perturbative, or we constrain them experimentally. 
As we do not rely on model-dependent values our findings are
more general than the explicit $U(1)_\text{FN} \times A_4 \times Z_3$ model under consideration.

Since the current LNU hints  in $R_D$ and in particular $R_{D^{*}}$ are too large to be accommodated with leptoquark flavor patterns,
there are also no joint explanations with $R_K$. 
 If both anomalies persist at the current level, additional BSM-physics would be required. 
 To have in both $R_K$ and $R_{D^{(*)}}$  LNU effects  maximized  from leptoquark flavor patterns requires  two types of  leptoquarks, with masses of at most at the level of several TeV. LFV signatures can be searched for with kaon decays and $\mu-e$ conversion.

\acknowledgments
We are happy to thank Stefan de Boer, Nejc Kosnik, Ivo de Medeiros Varzielas,  Ivan Nisandzic  and Erik Schumacher for useful discussions.
GH  is grateful to the Aspen Center for Physics where this project was finalized for its hospitality and stimulating environment.
The Aspen Center for Physics  is supported by National Science Foundation grant PHY-1066293.
This work is supported   by the  {\it  Bundesministerium f\"ur Bildung und Forschung} (BMBF).

\clearpage
\appendix
\section{Leptoquark couplings to SM fermions \label{app:L}}
\renewcommand{\arraystretch}{1.1}
\begin{table}[h]
    \centering
    \begin{tabular}{c|c|c}
        \hline
        $ \subset \mathcal{L}_\text{LQ} $  & $ ( SU(3)_C , SU(2)_L , Y ) $ & effective vertices \\
        \hline
        $\left(Y_{QL}\bar Q_\text L^{c} i\sigma_2L_\text{L} + Y_{UE}\bar u_\text{R}^c e_\text{R} \right) S_1^\dagger$ & (3,1,-1/3) & $\frac{Y_{QL}^{ij} \left( Y_{QL}^{mn} \right)^*}{2 M^2} ( \bar u_{\text L m} \gamma_\mu u_{\text L i} ) ( \bar \ell_{\text L n} \gamma^\mu \ell_{\text L j} )$\\ 
            &   & $-\frac{Y_{QL}^{ij} \left( Y_{QL}^{mn} \right)^*}{2 M^2} ( \bar u_{\text L m} \gamma_\mu d_{\text L i} ) ( \bar \ell_{\text L n} \gamma^\mu \nu_{\text L j} )$\\ 
            &   & $\frac{Y_{QL}^{ij} \left( Y_{QL}^{mn} \right)^*}{2 M^2} ( \bar d_{\text L m} \gamma_\mu d_{\text L i} ) ( \bar \nu_{\text L n} \gamma^\mu \nu_{\text L j} )$\\
            &   & $\frac{Y_{UE}^{ij} \left(Y_{UE}^{mn} \right)^*}{2 M^2} ( \bar u_{\text R m} \gamma_\mu u_{\text R i} ) ( \bar \ell_{\text R n} \gamma^\mu \ell_{\text R j} )$\\
            &   & $-\frac{Y_{QL}^{ij} \left( Y_{UE}^{mn} \right)^*}{2 M^2} ( \bar u_{\text R m} u_{\text L i} ) ( \bar \ell_{\text R n} \ell_{\text L j} )$\\
            &   & $\frac{Y_{QL}^{ij} \left( Y_{UE}^{mn} \right)^*}{8 M^2} ( \bar u_{\text R m} \sigma_{\mu\nu} u_{\text L i} ) ( \bar \ell_{\text R n} \sigma^{\mu\nu} \ell_{\text L j} )$\\
            &   & $\frac{Y_{QL}^{ij} \left( Y_{UE}^{mn} \right)^*}{2 M^2} ( \bar u_{\text R m} d_{\text L i} ) ( \bar \ell_{\text R n} \nu_{\text L j} )$\\
            &   & $-\frac{Y_{QL}^{ij} \left( Y_{UE}^{mn} \right)^*}{8 M^2} ( \bar u_{\text R m} \sigma_{\mu\nu} d_{\text L i} ) ( \bar \ell_{\text R n} \sigma^{\mu\nu} \nu_{\text L j} )$\\
        \hline 
	$ Y_{DE} \bar d_\text{R}^c e_\text R \tilde{S_1}^\dagger$ & (3,1,-4/3) & $\frac{Y_{DE}^{ij} \left( Y_{DE}^{mn} \right)^*}{2M^2} (\bar{d}_{\text{R}m} \gamma_\mu d_{\text{R}i}) (\bar{e}_{\text{R}n} \gamma^\mu e_{\text{R}j})$ \\ 
	\hline
	$\left( Y_{\bar{U}L} \bar u_\text{R} L_\text{L} + Y_{\bar{Q}E} \bar Q_\text{L} i \sigma_2 e_\text{R} \right)  S_2^\dagger$ & (3,2,-7/6) & $ - \frac{Y_{\bar{U}L}^{ij} \left( Y_{\bar{U}L}^{mn} \right)^* }{2M^2} (\bar u_{\text R i} \gamma_\mu u_{\text R m})(\bar \nu_ {\text L n} \gamma^\mu \nu_{\text L j} ) $\\
            &   & $ - \frac{Y_{\bar{U}L}^{ij} \left( Y_{\bar{U}L}^{mn} \right)^* }{2M^2} (\bar u_{\text R i} \gamma_\mu u_{\text R m})(\bar \ell_ {\text L n} \gamma^\mu \ell_{\text L j} ) $\\
            &   & $ - \frac{Y_{\bar{Q}E}^{ij} \left( Y_{\bar{Q}E}^{mn} \right)^* }{2M^2} (\bar u_{\text L i} \gamma_\mu u_{\text L m})(\bar \ell_ {\text R n} \gamma^\mu \ell_{\text R j} ) $\\
            &   & $ - \frac{Y_{\bar{Q}E}^{ij} \left( Y_{\bar{Q}E}^{mn} \right)^* }{2M^2} (\bar d_{\text L i} \gamma_\mu d_{\text L m})(\bar \ell_ {\text R n} \gamma^\mu \ell_{\text R j} ) $\\
            &   & $  \frac{Y_{\bar{U}L}^{ij} \left( Y_{\bar{Q}E}^{mn} \right)^* }{2M^2} (\bar u_{\text R i} d_{\text L m})(\bar \ell_ {\text R n} \nu_{\text L j} ) $\\
            &   & $  \frac{Y_{\bar{U}L}^{ij} \left( Y_{\bar{Q}E}^{mn} \right)^* }{8M^2} (\bar u_{\text R i} \sigma_{\mu\nu} d_{\text L m})(\bar \ell_ {\text R n} \sigma^{\mu\nu} \nu_{\text L j} ) $\\
	    &   & $ - \frac{Y_{\bar{U}L}^{ij} \left( Y_{\bar{Q}E}^{mn} \right)^* }{2M^2} (\bar u_{\text R i} u_{\text L m})(\bar \ell_ {\text R n} \ell_{\text L j} ) $\\
            &   & $ - \frac{Y_{\bar{U}L}^{ij} \left( Y_{\bar{Q}E}^{mn} \right)^* }{8M^2} (\bar u_{\text R i} \sigma_{\mu\nu} u_{\text L m})(\bar \ell_ {\text R n} \sigma^{\mu\nu} \ell_{\text L j} ) $\\
        \hline 
	$Y_{\bar{D}L} \bar d_\text{R} L_\text{L} \tilde{S}_2^\dagger$ & (3,2,1/6) & $-\frac{Y_{\bar{D}L}^{ij} \left( Y_{\bar{D}L}^{mn} \right)^*}{2M^2} (\bar{d}_{\text{R}i} \gamma_\mu d_{\text{R}m}) (\bar{\nu}_{\text{L}n} \gamma^\mu \nu_{\text{L}j})$ \\
	    &   & $-\frac{Y_{\bar{D}L}^{ij} \left( Y_{\bar{D}L}^{mn} \right)^*}{2M^2} (\bar{d}_{\text{R}i} \gamma_\mu d_{\text{R}m}) (\bar{\ell}_{\text{L}n} \gamma^\mu \ell_{\text{L}j})$ \\
	\hline
        $Y_{QL} \bar Q_\text{L}^c i \sigma_2 \vec{\sigma} L_\text{L} \vec{S}_3^\dagger$ & (3,3,-1/3) & $  \frac{ Y_{QL}^{ij} \left( Y_{QL}^{mn} \right)^* }{M^2} (\bar u_{\text L m} \gamma_\mu u_{\text L i}) (\bar \nu_{\text L n} \gamma^\mu \nu_{\text L j} ) $ \\
            &   & $  \frac{ Y_{QL}^{ij} \left( Y_{QL}^{mn} \right)^* }{M^2} (\bar d_{\text L m} \gamma_\mu d_{\text L i}) (\bar \ell_{\text L n} \gamma^\mu \ell_{\text L j} ) $ \\
            &   & $  \frac{ Y_{QL}^{ij} \left( Y_{QL}^{mn} \right)^* }{2M^2} (\bar u_{\text L m} \gamma_\mu u_{\text L i}) (\bar \ell_{\text L n} \gamma^\mu \ell_{\text L j} ) $ \\
            &   & $  \frac{ Y_{QL}^{ij} \left( Y_{QL}^{mn} \right)^* }{2M^2} (\bar u_{\text L m} \gamma_\mu d_{\text L i}) (\bar \ell_{\text L n} \gamma^\mu \nu_{\text L j} ) $ \\
            &   & $  \frac{ Y_{QL}^{ij} \left( Y_{QL}^{mn} \right)^* }{2M^2} (\bar d_{\text L m} \gamma_\mu d_{\text L i}) (\bar \nu_{\text L n} \gamma^\mu \nu_{\text L j} ) $ \\
        \hline
    \end{tabular}
    \caption{Scalar leptoquark models and their respective effective vertices at tree level. $Q$ denotes the $SU(2)_L$ doublet $(u_L \ d_L )$, with $u = u,c,t$; $d=d,s,b$; $\ell = e,\mu,\tau$ and $\nu = \nu_e, \nu_\mu, \nu_\tau$. Small roman indices are generation indices and are supressed in the first column. $Y=Q_e - I_3$ is the hypercharge, $Q_e$ the electric charge and $I_3$ the third component of the weak isospin.}
    \label{tab:effectivevertices_scalars}   
\end{table}

\renewcommand{\arraystretch}{1.4}
\begin{table}[h]
    \centering
    \begin{tabular}{c|c|c}
        \hline
        $ \subset \mathcal{L}_\text{LQ} $  & $ ( SU(3)_C , SU(2)_L , Y ) $ & effective vertices \\
        \hline
        $\left( Y_{\bar{Q}L} \bar Q_\text{L} \gamma_\mu L_\text{L} + Y_{\bar{D}E} \bar d_\text{R} \gamma_\mu e_\text{R} \right) V_1^{\mu \dagger}$ & (3,1,-1/6) & $-\frac{Y_{\bar{Q}L}^{ij} \left( Y_{\bar{Q}L}^{mn} \right)^*}{M^2} (\bar u_{\text L i} \gamma_\mu u_{\text L m}) (\bar \nu_{\text L n} \gamma^\mu \nu_{\text L j} ) $ \\
            &   & $-\frac{Y_{\bar{Q}L}^{ij} \left( Y_{\bar{Q}L}^{mn} \right)^*}{M^2} (\bar u_{\text L i} \gamma_\mu d_{\text L m}) (\bar \ell_{\text L n} \gamma^\mu \nu_{\text L j} ) $ \\
            &   & $-\frac{Y_{\bar{Q}L}^{ij} \left( Y_{\bar{Q}L}^{mn} \right)^*}{M^2} (\bar d_{\text L i} \gamma_\mu d_{\text L m}) (\bar \ell_{\text L n} \gamma^\mu \ell_{\text L j} ) $ \\
            &   & $-\frac{Y_{\bar{D}E}^{ij} \left( Y_{\bar{D}E}^{mn} \right)^*}{M^2} (\bar d_{\text R i} \gamma_\mu d_{\text R m}) (\bar \ell_{\text R n} \gamma^\mu \ell_{\text R j} ) $ \\
            &   & $ \frac{2 Y_{\bar{Q}L}^{ij} \left( Y_{\bar{D}E}^{mn} \right)^*}{M^2} (\bar u_{\text L i} d_{\text R m}) (\bar \ell_{\text R n} \nu_{\text L j} ) $ \\
            &   & $ \frac{2 Y_{\bar{Q}L}^{ij} \left( Y_{\bar{D}E}^{mn} \right)^*}{M^2} (\bar d_{\text L i} d_{\text R m}) (\bar \ell_{\text R n} \ell_{\text L j} ) $ \\
        \hline
	$Y_{\bar{U}E} \bar{u}_\text{R} \gamma_\mu e_\text{R} \tilde{V}_1^{\mu\dagger}$ & (3,1,5/3) & $-\frac{Y_{\bar{U}E}^{ij} \left( Y_{\bar{U}E}^{mn} \right)^*}{M^2} (\bar u_{\text R i} \gamma_\mu u_{\text{R}m}) (\bar \ell_{\text R n} \gamma^\mu \ell_{\text R j})$ \\
	\hline
        $\left( Y_{DL} \bar d_\text{R}^c \gamma_\mu L_\text{L} + Y_{QE} \bar Q_\text{L}^c \gamma_\mu e_\text{R} \right) i \sigma_2 V_2^{\mu \dagger}$ & (3,2,-5/6) & $\frac{Y_{DL}^{ij} \left( Y_{DL}^{mn} \right)^*}{M^2} (\bar d_{\text R m} \gamma_\mu d_{\text R i}) (\bar \nu_{\text L n} \gamma^\mu \nu_{\text L j} ) $ \\
            &   & $\frac{Y_{DL}^{ij} \left( Y_{DL}^{mn} \right)^*}{M^2} (\bar d_{\text R m} \gamma_\mu d_{\text R i}) (\bar \ell_{\text L n} \gamma^\mu \ell_{\text L j} ) $ \\
            &   & $\frac{Y_{QE}^{ij} \left( Y_{QE}^{mn} \right)^*}{M^2} (\bar u_{\text L m} \gamma_\mu u_{\text L i}) (\bar \ell_{\text R n} \gamma^\mu \ell_{\text R j} ) $ \\
            &   & $\frac{Y_{QE}^{ij} \left( Y_{QE}^{mn} \right)^*}{M^2} (\bar d_{\text L m} \gamma_\mu d_{\text L i}) (\bar \ell_{\text R n} \gamma^\mu \ell_{\text R j} ) $ \\
            &   & $\frac{2 Y_{DL}^{ij} \left( Y_{QE}^{mn} \right)^*}{M^2} (\bar u_{\text L m} d_{\text R i}) (\bar \ell_{\text R n} \nu_{\text L j} ) $ \\
            &   & $\frac{2 Y_{DL}^{ij} \left( Y_{QE}^{mn} \right)^*}{M^2} (\bar d_{\text L m} d_{\text R i}) (\bar \ell_{\text R n} \ell_{\text L j} ) $ \\
        \hline
	$Y_{UL} \bar{u}_\text{R}^c \gamma_\mu L_L \tilde{V}_2^{\mu\dagger}$ & (3,2,1/6) & $\frac{Y_{UL}^{ij} \left( Y_{UL}^{mn} \right)^*}{M^2} (\bar u_{\text R m} \gamma_\mu u_{\text{R}i}) (\bar \nu_{\text L n} \gamma^\mu \nu_{\text L j})$ \\
	    &   & $\frac{Y_{UL}^{ij} \left( Y_{UL}^{mn} \right)^*}{M^2} (\bar u_{\text R m} \gamma_\mu u_{\text{R}i}) (\bar \ell_{\text L n} \gamma^\mu \ell_{\text L j})$ \\
	\hline
        $ Y_{\bar{Q}L} \bar Q_\text{L} \gamma_\mu \vec{\sigma} L_\text{L} \vec{V}_3^{\mu\dagger}$ & (3,3,-2/3) & $-\frac{ 2 Y_{\bar{Q}L}^{ij} \left( Y_{\bar{Q}L}^{mn} \right)^*}{M^2} (\bar u_{\text L i} \gamma_\mu u_{\text L m}) (\bar \ell_{\text L n} \gamma^\mu \ell_{\text L j} ) $ \\
            &   & $-\frac{2 Y_{\bar{Q}L}^{ij} \left( Y_{\bar{Q}L}^{mn} \right)^*}{M^2} (\bar d_{\text L i} \gamma_\mu d_{\text L m}) (\bar \nu_{\text L n} \gamma^\mu \nu_{\text L j} ) $ \\
            &   & $-\frac{Y_{\bar{Q}L}^{ij} \left( Y_{\bar{Q}L}^{mn} \right)^*}{M^2} (\bar u_{\text L i} \gamma_\mu u_{\text L m}) (\bar \nu_{\text L n} \gamma^\mu \nu_{\text L j} ) $ \\
            &   & $ \frac{Y_{\bar{Q}L}^{ij} \left( Y_{\bar{Q}L}^{mn} \right)^*}{M^2} (\bar u_{\text L i} \gamma_\mu d_{\text L m}) (\bar \ell_{\text L n} \gamma^\mu \nu_{\text L j} ) $ \\
            &   & $-\frac{ Y_{\bar{Q}L}^{ij} \left( Y_{\bar{Q}L}^{mn} \right)^*}{M^2} (\bar d_{\text L i} \gamma_\mu d_{\text L m}) (\bar \ell_{\text L n} \gamma^\mu \ell_{\text L j} ) $ \\
        \hline
        \end{tabular}
    \caption{Same as Table \ref{tab:effectivevertices_scalars} but for vector leptoquark models.}
    \label{tab:effectivevertices_vectors}
\end{table}
\renewcommand{\arraystretch}{1}

\clearpage
\newpage

\section{\texorpdfstring{$b \to c \ell \nu$}{b to c l nu} \label{app:cc}}

The effective Hamiltonian for $b\to c\ell\nu$ transitions can be written as
\begin{equation}
	\mathcal H_\text{eff}^{b\to c\ell\nu} = \frac{4G_\text F}{\sqrt 2}V_{cb}\left(\delta_{\ell\nu}\mathcal O_{V_1}^{\ell\nu} + \sum_i C_i^{\ell\nu}\mathcal O_i^{\ell\nu}\right) \, , 
\end{equation}
where
\begin{equation}
	\begin{aligned}
		\mathcal O_{V_{1(2)}}^{\ell\nu} &= \left[\bar c_\text{L(R)}\gamma^\mu b_\text{L(R)}\right]\left[\bar\ell_\text L\gamma_\mu\nu_\text L\right]  \, ,\\
		\mathcal O_{S_{1(2)}}^{\ell\nu} &= \left[\bar c_\text{L(R)}b_\text{R(L)}\right]\left[\bar\ell_\text R\nu_\text L\right] \, , \\   
		\mathcal O_T^{\ell\nu} &= \left[\bar c_\text R\sigma^{\mu\nu}b_{\text L\vphantom(}\right]\left[\bar\ell_\text R\sigma_{\mu\nu}\nu_\text L\right] \, .
	\end{aligned}
\end{equation}
In the SM, all Wilson coefficients $C_i^{\ell\nu}$ vanish.
Tree level  contributions from leptoquarks are given in Table~\ref{tab:btoc_wilsons}, where $\tilde{C} \equiv  \frac{4G_\text F}{\sqrt 2}V_{cb} M^2 C$ is used for
brevity~\footnote{The chirality-flipping contribution given in Table 3 of \cite{Dorsner:2016wpm} for $S=0$ misses an overall sign. We thank Nejc Kosnik for clarification.}. Contributions to $\mathcal O_{V_{2}}^{\ell\nu} $ are  not induced at leading order.

The energy scale dependence of the Wilson coefficients is governed by renormalization group equations. At  leading logarithmic order holds \cite{Chetyrkin:1997dh,Gracey:2000am}
\begin{align}
	C_S (\mu_b) &= \left[ \frac{\alpha_s (m_t)}{\alpha_s (\mu_b)} \right]^\frac{\gamma_S}{2\beta_0^{(5)}} \left[ \frac{\alpha_s (M)}{\alpha_s (m_t)} \right]^\frac{\gamma_S}{2\beta_0^{(6)}} C_S (M)\,, &
	C_S (\mu_c) &= \left[ \frac{\alpha_s (\mu_b)}{\alpha_s (\mu_c)} \right]^\frac{\gamma_S}{2\beta_0^{(4)}} C_S (\mu_b)\,, \\
	C_T (\mu_b) &= \left[ \frac{\alpha_s (m_t)}{\alpha_s (\mu_b)} \right]^\frac{\gamma_T}{2\beta_0^{(5)}} \left[ \frac{\alpha_s (M)}{\alpha_s (m_t)} \right]^\frac{\gamma_T}{2\beta_0^{(6)}} C_T (M)\,, &
	C_T (\mu_c) &= \left[ \frac{\alpha_s (\mu_b)}{\alpha_s (\mu_c)} \right]^\frac{\gamma_T}{2\beta_0^{(4)}} C_T (\mu_b)
\end{align}
with 
\begin{align}
	\gamma_S &= -8\,, & \gamma_T &= \frac{8}{3}\,, \quad
	& \beta_0^{(n_f)} = 11 - \frac{2 n_f}{3}\,.
\end{align}  
We use the  {\tt CRunDec} package \cite{Schmidt:2012az} to evaluate $\alpha_s$. Assuming $M \sim 1 \text{TeV}$  we find
a modification of the Fierz relations between scalar and tensor operators, $ C_S (M) =  \mp  4 C_T (M)$, at the $b$- and $c$-quark mass scale, $\mu_b$ and $\mu_c$, respectively, 
\begin{align}
	C_S (\mu_b) &=  \mp \begin{cases} 7.8 C_T (\mu_b) \ M = \SI{1}{TeV} \\  8.2 C_T (\mu_b) \ M = \SI{2}{TeV} \\ 8.4 C_T (\mu_b)  \ M = \SI{3}{TeV} \end{cases} , \\
	C_S (\mu_c) &=  \mp \begin{cases} 11.0 C_T (\mu_c) \ M = \SI{1}{TeV}  \\ 11.6 C_T (\mu_c) \ M = \SI{2}{TeV} \\ 12.0 C_T (\mu_c) \ M = \SI{3}{TeV} \end{cases} \, , 
\end{align}
where the minus sign (plus sign) refers to scenario  $S_1$ ($S_2$).
The overall running of $C_S$ and $C_T$ is negligible compared to the unknown $\mathcal{O} (1)$ coefficients of the flavor patterns and will not be considered further.

The branching fractions of $B\to D^{(*)}\ell\nu$ decays can be written as
\begin{equation}  \label{eq:B8}
	\begin{split}
		\mathcal B(B\to D \ell \nu)=&  \sum\limits_\nu \mathcal B^\text{SM}(B\to D \ell \nu) |\delta_{\ell \nu} + C_{V_1}^{\ell \nu} + C_{V_2}^{\ell \nu}|^2 + A_{S} |C_{S_1}^{\ell \nu} + C_{S_2}^{\ell \nu}|^2 \\
		&+A_{T} |C_T^{\ell \nu}|^2 + A_{VS} \text{Re} \left[ (\delta_{\ell \nu} + C_{V_1}^{\ell \nu} + C_{V_2}^{\ell \nu})(C_{S_1}^{\ell \nu} + C_{S_2}^{\ell \nu})^* \right] \\
		&+  A_{VT} \text{Re} \left[ (\delta_{\ell \nu} + C_{V_1}^{\ell \nu} + C_{V_2}^{\ell \nu}) C_{T}^{\ell \nu*} \right]  \, , 
	\end{split}
\end{equation}
\begin{equation} \label{eq:B9}
	\begin{split}
		\mathcal B(B\to D^* \ell \nu)=& \sum\limits_\nu \mathcal B^\text{SM}(B\to D^* \ell \nu)\left[ |\delta_{\ell \nu} + C_{V_1}^{\ell \nu}|^2 + |C_{V_2}^{\ell \nu}|^2\right] \\
		& + B_{V_1V_2} \text{Re} \left[  ( \delta_{\ell \nu} + C_{V_1}^{\ell \nu} ) C_{V_2}^{\ell \nu *} \right] + B_{S} |C_{S_1}^{\ell \nu} - C_{S_2}^{\ell \nu}|^2 + B_{T} |C_T^{\ell \nu}|^2 \\
		&+ B_{VS} \text{Re} \left[ (\delta_{\ell \nu} + C_{V_1}^{\ell \nu} - C_{V_2}^{\ell \nu})(C_{S_1}^{\ell \nu} - C_{S_2}^{\ell \nu})^* \right] \\
		&+ B_{V_1 T} \text{Re} \left[ (\delta_{\ell \nu} + C_{V_1}^{\ell \nu} ) C_{T}^{\ell \nu*} \right] + B_{V_2 T} \text{Re} \left[ C_{V_2}^{\ell \nu} C_{T}^{\ell \nu *} \right] \, ,
	\end{split}
\end{equation}
where the coefficients $A_i$ and $B_i$ generally depend on the lepton and its polarization. Corresponding indices are
suppressed   in Eqs.~(\ref{eq:B8}), (\ref{eq:B9}) to avoid clutter.
The coefficients can be expressed in terms of hadronic matrix elements provided in Ref.~\cite{Sakaki:2013bfa}.
Using lattice data from \cite{Na:2015kha} for the $B\to D$ form factors and the HQET form factors from \cite{Sakaki:2013bfa} for $B\to D^*$, we find the numerical values given in Tables \ref{tab:BtoD_coeff} and \ref{tab:BtoDstar_coeff} by integrating over the whole $q^2$-range, summing over the lepton-polarization and normalizing to the SM branching ratios.
For the latter  we obtain \mbox{$\mathcal B^\text{SM}(B^0\to D^+\tau\nu)=\num{6.66\pm 0.67e-3}$}, \mbox{$\mathcal B^\text{SM}(B^0\to D^+(e,\mu)\nu)=\num{2.23\pm 0.24e-2}$} and $\mathcal B^\text{SM}(B^0\to D^{+*}\tau\nu)=\num{1.35\pm 0.10e-2}$, $\mathcal B^\text{SM}(B^0\to D^{+*}(e,\mu)\nu)=\num{5.34\pm 0.40e-2}$.
Here we use the lifetime $\tau_{B^0}=\SI{1.520\pm0.004e-12}{s}$ of the $B^0$ meson \cite{Agashe:2014kda}.
In order to estimate the uncertainties, we draw \num{e5} random samples of the form factor parameters provided in the respective references and calculate the coefficients $A_i$ and $B_i$ for each sample.
The mean and standard deviation of the resulting distributions are then considered as the central value and uncertainty.
We assume that the form factor parameters are normally distributed and incorperate all correlations provided by  \cite{Na:2015kha,Sakaki:2013bfa}.

Additionally, we provide the coefficients for given $\tau$-polarizations in Tables \ref{tab:BtoD_polarized_coeff} and \ref{tab:BtoDstar_polarized_coeff}, where we normalize to the difference $\mathcal B^\text{SM}_{k=+} - \mathcal B^\text{SM}_{k=-}$ of the SM values of the polarized branching fractions. For the latter we obtain  $\mathcal B^\text{SM}_{k=+}(B^0\to D^+\tau\nu)=\num{4.43\pm 0.47e-3}$, $\mathcal B^\text{SM}_{k=-}(B^0\to D^+\tau\nu)=\num{2.22\pm 0.22e-3}$, and $\mathcal B^\text{SM}_{k=+}(B^0\to D^{+*}\tau\nu)=\num{3.40\pm 0.27e-3}$, $\mathcal B^\text{SM}_{k=-}(B^0\to D^{+*}\tau\nu)=\num{1.01\pm 0.07e-2}$.
Scalar operators do not contribute to the case where $k=-$.
\begin{table}
	\centering
	\begin{tabular}{
			c||
			S[table-format=3.2(4)]|
			S[table-format=3.2(4)]|
			S[table-format=3.2(4)]|
			S[table-format=3.2(4)]
		}
		$\ \ell\ $ & {$\hat A_S^\ell$} & {$\hat A_T^\ell$} & {$\hat A_{VS}^\ell$} & {$\hat A_{VT}^\ell$} \\
		\hline
		$e$ & 1.45 \pm 0.16 & 0.38 \pm 0.20 & 0.00 \pm 0.00 & 0.00 \pm 0.00 \\
		$\mu$ & 1.45 \pm 0.16 & 0.36 \pm 0.17 & 0.17 \pm 0.02 & 0.13 \pm 0.09 \\
		$\tau$ & 1.36 \pm 0.15 & 0.35 \pm 0.13 & 1.73 \pm 0.19 & 0.69 \pm 0.15 \\
	\end{tabular}
	\caption{The normalized $B\to D\ell\nu$ coefficients  $\hat A_i^\ell = A_i^\ell/\mathcal B^\text{SM}$.}
	\label{tab:BtoD_coeff}
\end{table}
\begin{table}
	\centering
	\begin{tabular}{
			c||
			S[table-format=3.2(4)]|
			S[table-format=3.2(4)]|
			S[table-format=3.2(4)]|
			S[table-format=3.2(4)]|
			S[table-format=3.2(4)]|
			S[table-format=3.2(4)]
		}
		$\ \ell\ $ & {$\hat B_{V_1V_2}^\ell$} & {$\hat B_S^\ell$} & {$\hat B_T^\ell$} & {$\hat B_{VS}^\ell$} & {$\hat B_{V_1T}^\ell$} & {$\hat B_{V_2T}^\ell$} \\
		\hline
		$e$ & -1.72 \pm 0.13 & 0.06 \pm 0.01 & 12.98 \pm 0.98 & 0.00 \pm 0.00 & 0.00 \pm 0.00 & 0.00 \pm 0.00 \\
		$\mu$ & -1.72 \pm 0.13 & 0.06 \pm 0.01 & 12.98 \pm 0.98 & 0.02 \pm 0.00 & -0.43 \pm 0.03 & 0.70 \pm 0.05 \\
		$\tau$ & -1.78 \pm 0.13 & 0.04 \pm 0.01 & 13.35 \pm 1.00 & 0.12 \pm 0.01 & -4.58 \pm 0.34 & 6.14 \pm 0.45 \\
	\end{tabular}
	\caption{The normalized $B\to D^*\ell\nu$ coefficients $\hat B_i^\ell = B_i^\ell/\mathcal B^\text{SM}$.}
	\label{tab:BtoDstar_coeff}
\end{table}
\begin{table}
	\centering
	\begin{tabular}{
			c||
			S[table-format=3.2(4)]|
			S[table-format=3.2(4)]|
			S[table-format=3.2(4)]|
			S[table-format=3.2(4)]
		}
		$\ k\ $ & {$\hat A_S^k$} & {$\hat A_T^k$} & {$\hat A_{VS}^k$} & {$\hat A_{VT}^k$} \\
		\hline
		$+$ & 4.12 \pm 0.45 & 0.56 \pm 0.20 & 5.23 \pm 0.57 & 0.70 \pm 0.15 \\
		$-$ & {-} & 0.50 \pm 0.19 & {-} & 1.39 \pm 0.30 \\

	\end{tabular}
	\caption{The normalized $B\to D\tau \nu$ coefficients $\hat A_i^k = A_i^k/\left(\mathcal B^\text{SM}_{k=+}~-~\mathcal B^\text{SM}_{k=-}\right)$ for a given polarization $k$ of the $\tau$ lepton.}
	\label{tab:BtoD_polarized_coeff}
\end{table}
\begin{table}
	\centering
	\begin{tabular}{
			c||
			S[table-format=3.2(4)]|
			S[table-format=3.2(4)]|
			S[table-format=4.2(4)]|
			S[table-format=3.2(4)]|
			S[table-format=3.2(4)]|
			S[table-format=3.2(4)]
		}
		$\ k\ $ & {$\hat B_{V_1V_2}^k$} & {$\hat B_S^k$} & {$\hat B_T^k$} & {$\hat B_{VS}^k$} & {$\hat B_{V_1T}^k$} & {$\hat B_{V_2T}^k$} \\
		\hline
		$+$ & 0.62 \pm 0.06 & 0.04 \pm 0.00 & -14.15 \pm 1.06 & -0.24 \pm 0.03 & 3.08 \pm 0.23 & -4.12 \pm 0.30 \\
		$-$ & 1.26 \pm 0.10 & {-} & -12.72 \pm 0.95 & {-} & 6.15 \pm 0.46 & -8.24 \pm 0.61 \\

	\end{tabular}
	\caption{The normalized $B\to D^*\tau \nu$ coefficients $\hat B_i^k = B_i^k/\left(\mathcal B^\text{SM}_{k=+}~-~\mathcal B^\text{SM}_{k=-}\right)$ for a given polarization $k$ of the $\tau$ lepton.}
	\label{tab:BtoDstar_polarized_coeff}
\end{table}
\begin{table}
    \centering
    \begin{tabular}{c||c|c|c|c}
            & $\tilde{C}_{V_1}$ & $\tilde{C}_{S_1}$ & $\tilde{C}_{S_2}$ & $\tilde{C}_{T}$ \\ \hline
        $S_1$         & $\frac{1}{2} Y_{QL}^{b\nu} \left( Y_{QL}^{c\ell} \right)^*$ & - & $-\frac{1}{2} Y_{QL}^{b\nu} \left( Y_{UE}^{c\ell} \right)^*$ & $\frac{1}{8} Y_{QL}^{b\nu} \left( Y_{UE}^{c\ell} \right)^*$  \\
        $\tilde{S}_1$ & - & - & - & - \\
        $S_2$         & - & - & $-\frac{1}{2} Y_{\bar{U}L}^{c\nu} \left( Y_{\bar{Q}E}^{b\ell} \right)^*$ & $-\frac{1}{8} Y_{\bar{U}L}^{c\nu} \left( Y_{\bar{Q}E}^{b\ell} \right)^*$\\
        $\tilde{S}_2$ & - & - & - & - \\
        $S_3$         & $- \frac{1}{2} Y_{QL}^{b\nu} \left( Y_{QL}^{c\ell} \right)^*$ & - & - & - \\
        $V_1$         & $Y_{\bar{Q}L}^{c\nu} \left( Y_{\bar{Q}L}^{b\ell} \right)^*$ & $- 2 Y_{\bar{Q}L}^{c\nu} \left( Y_{\bar{D}E}^{b\ell} \right)^*$ & - & - \\
        $\tilde{V}_1$ & - & - & - & - \\
        $V_2$         & - & $- 2 Y_{DL}^{b\nu} \left( Y_{QE}^{c\ell} \right)^*$ & - & - \\
        $\tilde{V}_2$ & - & - & - & - \\
        $V_3$ & $- Y_{\bar{Q}L}^{c\nu} \left( Y_{\bar{Q}L}^{b\ell} \right)^*$ & - & - & - \\ 
    \end{tabular}
    \caption{Contributions from leptoquarks to $b\to c \ell \nu$ transitions at matching scale.}
    \label{tab:btoc_wilsons}
\end{table} 
\FloatBarrier

\section{\texorpdfstring{$b \to s \ell \ell'$ and $b \to s \nu \nu'$}{b to s ll and b to s nu nu}  \label{app:bsll}}

The $b\to s\ell\ell'$ and $b\to s\nu\nu'$ processes can be described by the effective Hamiltionian
\begin{equation}
	\mathcal H_\text{eff}^{b\to s\ell\ell'(\nu\nu')} = -\frac{4G_\text F}{\sqrt 2}V_{tb}V_{ts}^*\frac{\alpha}{4\pi}\sum_iC_i\mathcal O_i \, , 
\end{equation}
with the effective operators
\begin{equation}
	\begin{aligned}
		\mathcal O_9^{(\prime)\ell\ell'} &= \left[\bar s\gamma_\mu P_\text{L(R)}b\right]\left[\bar \ell'\gamma^\mu\ell\right]\,, & \mathcal O_{10}^{(\prime)\ell\ell'} &= \left[\bar s\gamma_\mu P_\text{L(R)}b\right]\left[\bar \ell'\gamma^\mu\gamma_5\ell\right]\,, \\
		\mathcal O_\text{S}^{(\prime)\ell\ell'} &= [\bar sP_\text{R(L)}b][\bar\ell'\ell]\,, & \mathcal O_\text{P}^{(\prime)\ell\ell'} &= [\bar sP_\text{R(L)}b][\bar\ell'\gamma_5\ell]\,, \\
		\mathcal O_\text{L(R)}^{\nu\nu'} &= \left[\bar s\gamma_\mu P_\text{L(R)}b\right]\left[\bar \nu'\gamma^\mu P_\text L\nu\right]\,, &&
	\end{aligned}
\end{equation}
which, in general, depend on the flavor of the  leptons.
In the SM, the relevant Wilson coefficients for $b\to s\ell\ell$ transitions are $C_9^\text{SM}\simeq -C_{10}^\text{SM}\simeq 4.2$ at the $m_b$ scale, universally for all leptons, while contributions to the scalar operators are negligible.
Table \ref{tab:btos_wilsons} shows the leptoquark  tree level contributions to the Wilson coefficients, where we use $\tilde C \equiv \frac{4G_\text F}{\sqrt 2}V_{tb}V_{ts}^*\frac{\alpha}{4\pi}M^2 C$ for brevity.
For $b \to s \nu \nu$, 
$C_\text L^\text{SM} = -\frac{2X_t}{\sin^2\theta_\text W} \simeq -13$ and $C_\text R^\text{SM}$ is negligible.
The strongest bound on new physics (NP) in $b\to s\nu\nu$ transitions is provided by  $\mathcal B(B^+\to K^+\nu\nu) < \num{1.7e-5} \text{ at } \SI{90}{\percent}\text{\,CL}$ \cite{Lees:2013kla}, which, using \cite{Buchalla:2000sk}, implies an enhancement over the SM of at most a factor of $4.3$. Therefore,
\begin{equation}
	\sqrt{  \sum_\nu  |C_\text L^{\rm SM} + C_\text L^{ {\rm NP} \nu \nu}+ C^{\nu \nu}_\text R|^2 +\sum_{\nu \neq \nu^\prime} | C^{\nu \nu^\prime}_\text L + C^{\nu \nu^\prime }_\text R  |^2}  \leq   |C_\text L^{\rm SM}|  \sqrt{4.3 \cdot  3}\simeq  47 \, .
	\label{eq:BtoKnunu_bound_L}
\end{equation}
Solving this for a single, dominant diagonal coupling $C_\text L^{ {\rm NP}  \nu \nu}$ gives
$-30 \leq C_\text L^{ {\rm NP}\nu \nu}\leq 56$. This implies constraints on 
 leptoquark yukawas to third generation lepton in models $S_1,V_3$ with Fierz factors $m(\Delta)=1/2,-2$, respectively, through  the relation
\begin{align}
YY^*|_\tau =\frac{1}{m(\Delta)} 
\frac{4 G_F}{\sqrt{2}} V_{tb}V_{ts}^*\frac{\alpha}{4\pi}   C_\text L^{ {\rm NP}\nu_\tau \nu_\tau} M^2 \simeq 7.9 \cdot 10^{-4} \,  \frac{   C_\text L^{ {\rm NP}\nu_\tau \nu_\tau}   }{m(\Delta)} 
\left(  \frac{M}{\mbox{TeV}}\right)^2  \, .
\end{align}
For $V_3$,  $YY^*|_\tau \lesssim 0.02 (M/{\rm TeV})^2$.

\clearpage
\begin{turnpage}
	\begin{table}
		\centering
		\begin{tabular}{c||c|c|c|c|c|c|c|c|c|c}
			& $\tilde C_9$ & $\tilde C_{10}$ & $\tilde C_9'$ & $\tilde C_{10}'$ & $\tilde C_\text S$ & $\tilde C_\text P$ & $\tilde C_\text S'$ & $\tilde C_\text P'$ & $\tilde C_\text L$ & $\tilde C_\text R$ \\ \hline
			$S_1$ & - & - & - & - & - & - & - & - & $\frac12Y_{QL}^{b\ell}\left(Y_{QL}^{s\ell'}\right)^*$ & - \\
			$\tilde S_1$ & - & - & $\frac14Y_{DE}^{b\ell}\left(Y_{DE}^{s\ell'}\right)^*$ & $+\tilde C_9'$ & - & - & - & - & - & - \\
			$S_2$ & $-\frac14Y_{\bar QE}^{s\ell}\left(Y_{\bar QE}^{b\ell'}\right)^*$ & $+\tilde C_9$ & - & - & - & - & - & - & - & - \\
			$\tilde S_2$ & - & - & $-\frac14Y_{\bar DL}^{s\ell}\left(Y_{\bar DL}^{b\ell'}\right)^*$ & $-\tilde C_9'$ & - & - & - & - & - & $-\frac12Y_{\bar DL}^{s\ell}\left(Y_{\bar DL}^{b\ell'}\right)^*$ \\
			$S_3$ & $\frac12Y_{QL}^{b\ell}\left(Y_{QL}^{s\ell'}\right)^*$ & $-\tilde C_9$ & - & - & - & - & - & - & $\frac12Y_{QL}^{b\ell}\left(Y_{QL}^{s\ell'}\right)^*$ & - \\
			$V_1$ & $- \frac12Y_{\bar QL}^{s\ell}\left(Y_{\bar QL}^{b\ell'}\right)^*$ & $-\tilde C_9$ & $- \frac12Y_{\bar DE}^{s\ell}\left(Y_{\bar DE}^{b\ell'}\right)^*$ & $+\tilde C_9'$ & $Y_{\bar QL}^{s\ell}\left(Y_{\bar DE}^{b\ell'}\right)^*$ & $\tilde C_\text S$ & $Y_{\bar DE}^{s\ell}\left(Y_{\bar QL}^{b\ell'}\right)^*$ & $+\tilde C_\text S'$ & - & - \\
			$\tilde V_1$ & - & - & - & - & - & - & - & - & - & - \\
			$V_2$ & - & - & $\frac12Y_{DL}^{b\ell}\left(Y_{DL}^{s\ell'}\right)^*$ & $-\tilde C_9'$ & $Y_{DL}^{b\ell}\left(Y_{QE}^{s\ell'}\right)^*$ & $-\tilde C_\text S$ & $Y_{QE}^{b\ell}\left(Y_{DL}^{s\ell'}\right)^*$ & $+\tilde C_\text S'$ & - & $Y_{DL}^{b\ell}\left(Y_{DL}^{s\ell'}\right)^*$ \\
			$\tilde V_2$ & - & - & - & - & - & - & - & - & - & - \\
			$V_3$ & $- \frac12Y_{\bar QL}^{s\ell}\left(Y_{\bar QL}^{b\ell'}\right)^*$ & $-\tilde C_9$ & - & - & - & - & - & - & $- 2Y_{\bar QL}^{s\ell}\left(Y_{\bar QL}^{b\ell'}\right)^*$ & - \\
		\end{tabular}
		\caption{Contributions from leptoquarks to $b\to s\ell\ell$ and $b\to s\nu\nu$ transitions at matching scale.}
		\label{tab:btos_wilsons}
	\end{table} 
\end{turnpage}
\FloatBarrier

\section{\texorpdfstring{$c \to u \ell \ell'$ and $c \to u \nu \nu'$}{c to u ll' and c to u nu nu'} \label{sec:cu}}

In order to describe the up-type FCNCs $c\to u\ell\ell'$ and $c\to u\nu\nu'$, we employ the effective Hamiltonian
\begin{equation}
	\mathcal H_\text{eff}^{c\to u\ell\ell'(\nu\nu')} = -\frac{4G_\text F}{\sqrt 2}\frac{\alpha}{4\pi}\sum_i\mathcal C_iQ_i,
\end{equation}
with the effective operators $Q_i$ defined as
\begin{equation}
	\begin{aligned}
		Q_9^{(\prime)\ell\ell'} &= \left[\bar u\gamma_\mu P_\text{L(R)}c\right]\left[\bar \ell'\gamma^\mu\ell\right]\,, & Q_{10}^{(\prime)\ell\ell'} &= \left[\bar u\gamma_\mu P_\text{L(R)}c\right]\left[\bar \ell'\gamma^\mu\gamma_5\ell\right]\,, \\
		Q_\text{S}^{(\prime)\ell\ell'} &= [\bar uP_\text{R(L)}c][\bar\ell'\ell]\,, & Q_\text{P}^{(\prime)\ell\ell'} &= [\bar uP_\text{R(L)}c][\bar\ell'\gamma_5\ell]\,, \\
		Q_\text T &= \left[\bar u\sigma_{\mu\nu} c\right]\left[\bar \ell'\sigma^{\mu\nu}\ell\right]\,, & Q_\text{T5} &= \left[\bar u\sigma_{\mu\nu} c\right]\left[\bar \ell'\sigma^{\mu\nu}\gamma_5\ell\right]\,, \\
		Q_\text{L(R)}^{\nu\nu'} &= \left[\bar u\gamma_\mu P_\text{L(R)}c\right]\left[\bar \nu'\gamma^\mu P_\text L\nu\right]\,. &&
	\end{aligned}
	\label{eq:btou_operators}
\end{equation}
In the SM the Wilson coefficients of these operators are small and can be neglected compared to the NP contributions \cite{deBoer:2015boa}.
We provide the leptoquark-induced contributions in Table \ref{tab:ctou_wilsons} using the shortcut notation $\tilde{\mathcal C} \equiv \frac{4G_\text F}{\sqrt 2}\frac{\alpha}{4\pi}M^2\mathcal C$.
For the sake of simplicity we introduce the coefficients $\mathcal C_{\mathrm T_{1,2}}$ which are related to those associated to the operator basis \eqref{eq:btou_operators} by $\tilde{\mathcal C}_\mathrm{T(5)} = \tilde{\mathcal C}_{\mathrm T_1} \pm \tilde{\mathcal C}_{\mathrm T_2}$.

\clearpage
\begin{turnpage}
	\begin{table}
		\centering
		\begin{tabular}{c||c|c|c|c|c|c|c|c|c|c|c|c}
			& $\tilde{\mathcal C}_9$ & $\tilde{\mathcal C}_{10}$ & $\tilde{\mathcal C}_9'$ & $\tilde{\mathcal C}_{10}'$ & $\tilde{\mathcal C}_\text S$ & $\tilde{\mathcal C}_\text P$ & $\tilde{\mathcal C}_\text S'$ & $\tilde{\mathcal C}_\text P'$ & $\tilde{\mathcal C}_{\mathrm T_1}$ & $\tilde{\mathcal C}_{\mathrm T_2}$ &$\tilde{\mathcal C}_\text L$ & $\tilde{\mathcal C}_\text R$ \\ \hline
			$S_1$ & $\frac14 Y_{QL}^{c\ell}\left(Y_{QL}^{u\ell'}\right)^*$ & $-\tilde{\mathcal C}_9$ & $\frac14 Y_{UE}^{c\ell}\left(Y_{UE}^{u\ell'}\right)^*$ & $+\tilde{\mathcal C}_9'$ & $-\frac14 Y_{UE}^{c\ell}\left(Y_{QL}^{u\ell'}\right)^*$ & $+\tilde{\mathcal C}_\text S$ & $-\frac14 Y_{QL}^{c\ell}\left(Y_{UE}^{u\ell'}\right)^*$ & $-\tilde{\mathcal C}_\text S'$ & $\frac18 Y_{UE}^{c\ell}\left(Y_{QL}^{u\ell'}\right)^*$ & $\frac18 Y_{QL}^{c\ell}\left(Y_{UE}^{u\ell'}\right)^*$ & - & - \\
			$\tilde S_1$ & -  & -  & -  & -  & -  & -  & -  & -  & -  & -  & -  & -  \\
			$S_2$ & $-\frac14 Y_{\bar QE}^{u\ell}\left(Y_{\bar QE}^{c\ell'}\right)^*$ & $+\tilde{\mathcal C}_9$ & $-\frac14 Y_{\bar UL}^{u\ell}\left(Y_{\bar UL}^{c\ell'}\right)^*$ & $-\tilde{\mathcal C}_9'$ & $\frac14 Y_{\bar QE}^{u\ell}\left(Y_{\bar UL}^{c\ell'}\right)^*$ & $+\tilde{\mathcal C}_\text S$ & $\frac14 Y_{\bar UL}^{u\ell}\left(Y_{\bar QE}^{c\ell'}\right)^*$ & $-\tilde{\mathcal C}_\text S'$ & $\frac18 Y_{\bar UL}^{u\ell}\left(Y_{\bar QE}^{c\ell'}\right)^*$ & $\frac18 Y_{\bar QE}^{u\ell}\left(Y_{\bar UL}^{c\ell'}\right)^*$ & - & $-\frac12 Y_{\bar UL}^{u\ell}\left(Y_{\bar UL}^{c\ell'}\right)^*$ \\
			$\tilde S_2$ & -  & -  & -  & -  & -  & -  & -  & -  & -  & -  & -  & -  \\
			$S_3$ & $\frac14 Y_{QL}^{c\ell}\left(Y_{QL}^{u\ell'}\right)^*$ & $-\tilde{\mathcal C}_9$ & - & - & - & - & - & - & - & - & $ Y_{QL}^{c\ell}\left(Y_{QL}^{u\ell'}\right)^*$ & - \\
			$V_1$ & -  & -  & -  & -  & -  & -  & -  & -  & -  & -  & $- Y_{\bar QL}^{u\ell}\left(Y_{\bar QL}^{c\ell'}\right)^*$ & - \\
			$\tilde V_1$ & - & - & $- \frac12 Y_{\bar UE}^{u\ell}\left(Y_{\bar UE}^{c\ell'}\right)^*$ & $+\tilde{\mathcal C}_9'$ & - & - & - & - & - & - & - & - \\
			$V_2$ & $\frac12 Y_{QE}^{c\ell}\left(Y_{QE}^{u\ell'}\right)^*$ & $+\tilde{\mathcal C}_9$ & - & - & - & - & - & - & - & - & - & - \\
			$\tilde V_2$ & - & - & $\frac12 Y_{UL}^{c\ell}\left(Y_{UL}^{u\ell'}\right)^*$ & $-\tilde{\mathcal C}_9'$ & - & - & - & - & - & - & - & $Y_{UL}^{c\ell}\left(Y_{UL}^{u\ell'}\right)^*$ \\
			$V_3$ & $- Y_{\bar QL}^{u\ell}\left(Y_{\bar QL}^{c\ell'}\right)^*$ & $-\tilde{\mathcal C}_9$ & - & - & - & - & - & - & - & - & $- Y_{\bar QL}^{u\ell}\left(Y_{\bar QL}^{c\ell'}\right)^*$ & - \\
		\end{tabular}
		\caption{Contributions from leptoquarks to $c\to u\ell\ell$ and $c\to u\nu\nu$ transitions at matching scale.}
		\label{tab:ctou_wilsons}
	\end{table} 
\end{turnpage}
\FloatBarrier

\end{document}